\documentclass[12pt,a4paper]{article}

\usepackage{amsmath}
\usepackage{fullpage}
\usepackage{url}
\usepackage{enumitem}
\usepackage{mathrsfs} 
\usepackage[dvips]{graphicx}
\usepackage{pst-plot}
\usepackage{pst-math}
\usepackage{pstricks}
\usepackage{pst-3dplot}
\usepackage{multido}
\usepackage{fp}
\usepackage{ifthen}

\newcommand{\Scri}{\mathscr I}
\newcommand{\Scrh}{\mathscr H}
\newcommand{\Scrs}{\mathscr S}

\usepackage{xcolor}
  \definecolor{forest}{rgb}{0,0.5,0}
  \definecolor{violet}{rgb}{0.33,0,0.67}
  \definecolor{Green}{rgb}{0,0.8,0.1} 

\newcommand{\rI}{Region~I}
\newcommand{\rII}{Region~II}
\newcommand{\rIII}{Region~III}
\newcommand{\rIV}{Region~IV}

\newcommand{\Schw}{Schwarzschild}
\newcommand{\KrSz}{Kruskal--Szekeres}

\newcommand{\tr}{\textrm Tr\,}

\newcommand{\rU}{\frac{\partial r}{\partial U}}
\newcommand{\rV}{\frac{\partial r}{\partial V}}
\newcommand{\fU}{\frac{\partial f}{\partial U}}
\newcommand{\fV}{\frac{\partial f}{\partial V}}

\newcommand{\p}{\partial}
\newcommand{\pb}[2]{		
   \parbox[t]{#1}{
      \raggedright
      #2
   }
}

\newcommand{\showlabel}[1]{%
\label{#1}%
}

\def\RI{			
  \FPmul{\xbr}{1}{\sx}		
  \FPmul{\ybr}{-1}{\sy}
  \FPmul{\xtr}{1}{\sx}		
  \FPmul{\ytr}{1}{\sy}
  \FPmul{\xc}{0}{\sx}		
  \FPmul{\yc}{0}{\sy}
  \FPmul{\xmr}{2}{\sx}		
  \FPmul{\ymr}{0}{\sy}
  \psline[linecolor=Green](\xbr,\ybr)(\xmr,\ymr)(\xtr,\ytr)
  \FPmul{\xI}{1.6}{\sx}
  \FPmul{\yI}{0.6}{\sy}
  \rput{-\ag}(\xI,\yI){$\Scri^+$}
  \FPmul{\xI}{1.6}{\sx}
  \FPmul{\yI}{-0.6}{\sy}
  \rput{-\ag}(\xI,\yI){$\Scri^-$}
  \psline[linecolor=red,linestyle=dashed](\xc,\yc)(\xbr,\ybr)
  \psline[linecolor=red,linestyle=dashed](\xc,\yc)(\xtr,\ytr)
  \FPmul{\xH}{0.6}{\sx}
  \FPmul{\yH}{0.4}{\sy}
  \rput{-\ag}(\xH,\yH){$\Scrh^+$}
  \FPmul{\xH}{0.62}{\sx}
  \FPmul{\yH}{-0.38}{\sy}
  \rput{-\ag}(\xH,\yH){$\Scrh^-$}
}

\def\RII{			
  \FPmul{\xtr}{1}{\sx}		
  \FPmul{\ytr}{1}{\sy}
  \FPmul{\xtl}{-1}{\sx}		
  \FPmul{\ytl}{1}{\sy}
  \FPmul{\xc}{0}{\sx}		
  \FPmul{\yc}{0}{\sy}
  \psline[linecolor=blue](\xtl,\ytl)(\xtr,\ytr)	
  \FPmul{\xS}{0}{\sx}
  \FPmul{\yS}{1.1}{\sy}
  \ifthenelse{\sy=1}
    {\rput(\xS,\yS){$\Scrs^+$}}
    {\rput(\xS,\yS){$\Scrs^-$}}
  \psline[linecolor=red,linestyle=dashed](\xc,\yc)(\xtr,\ytr)
  \psline[linecolor=red,linestyle=dashed](\xc,\yc)(\xtl,\ytl)
}

\def\UVarrows{
  \definecolor{Ppl}{rgb}{0.6,0,0.6}
  \FPmul{\xo}{-0.1}{\su}
  \FPmul{\yo}{-0.1}{\sv}
  \FPmul{\xa}{0.1}{\su}
  \FPmul{\ya}{0.1}{\sv}
  \FPmul{\xl}{0.18}{\su}
  \FPmul{\yl}{0.18}{\sv}
  \FPmul{\xm}{0.15}{\su}
  \FPmul{\ym}{0.15}{\sv}
  \rput{45}(0,0){
    \psline[linecolor=Ppl]{->}(\xo,\yo)(\xa,\yo)
    \rput{-45}(\xl,\yo){$V$}
    \psline[linecolor=Ppl]{->}(\xo,\yo)(\xo,\ya)
    \rput{-45}(\xo,\yl){$U$}
    \psline[linecolor=Ppl]{->}(\xo,\yo)(\xa,\ya)
    \rput{-45}(\xm,\ym){$T$}
  }
}

\def\trarrows{
  \definecolor{DkBl}{rgb}{0,0,0.7}
  \FPmul{\xp}{-0.15}{\sx}
  \FPmul{\xq}{0.15}{\sx}
  \FPset{\xO}{-0.4}
  \FPset{\ai}{-27}
  \FPset{\af}{27}
  \ifthenelse{\sy=1}
    {\psarc[linecolor=DkBl]{->}(\xO,0){-\xO}{\ai}{\af}}
    {\psarc[linecolor=DkBl]{<-}(\xO,0){-\xO}{\ai}{\af}}
  \FPmul{\xt}{-0.05}{1}
  \FPmul{\yt}{0.25}{\sy}
  \rput{-\ag}(\xt,\yt){$t$}
  \psline[linecolor=DkBl]{->}(\xp,0)(\xq,0)
  \FPmul{\xr}{0.21}{\sx}
  \rput{-\ag}(\xr,0){$r$}
}

\def\imzp{
  \FPmul{\xim}{1.1}{\sx}
  \FPmul{\yim}{-1.03}{\sy}
  \rput{-\ag}(\xim,\yim){$i^-$}
  \FPmul{\xiz}{2.1}{\sx}
  \FPmul{\yiz}{0}{\sy}
  \rput{-\ag}(\xiz,\yiz){$i^0$}
  \FPmul{\xip}{1.1}{\sx}
  \FPmul{\yip}{1.03}{\sy}
  \rput{-\ag}(\xip,\yip){$i^+$}
}

\def\TLcurve{           
  \FPmul{\xcva}{0.75}{\sx}
  \FPmul{\ycva}{-0.75}{\sy}
  \FPmul{\xcvb}{0.85}{\sx}
  \FPmul{\ycvb}{0}{\sy}
  \FPmul{\yarb}{-0.001}{\sy}
  \FPmul{\yart}{0.001}{\sy}
  \FPmul{\xcvc}{0.75}{\sx}
  \FPmul{\ycvc}{0.75}{\sy}
  \pscurve[linecolor=cyan](\xcva,\ycva)(\xcvb,\ycvb)(\xcvc,\ycvc)
  \psline[linecolor=cyan]{->}(\xcvb,\yarb)(\xcvb,\yart)
}

\def\SLcurve{           
  \FPmul{\xcva}{0.75}{\sx}
  \FPmul{\ycva}{-0.75}{\sy}
  \FPmul{\xcvb}{1.3}{\sx}
  \FPmul{\ycvb}{-0.6}{\sy}
  \FPmul{\xarb}{0.9}{\sx}
  \FPmul{\yarb}{-0.723}{\sy}
  \FPmul{\xart}{0.91}{\sx}
  \FPmul{\yart}{-0.721}{\sy}
  \FPmul{\xcvc}{2}{\sx}
  \FPmul{\ycvc}{0}{\sy}
  \pscurve[linecolor=cyan](\xcva,\ycva)(\xcvb,\ycvb)(\xcvc,\ycvc)
  \psline[linecolor=cyan]{->}(\xarb,\yarb)(\xart,\yart)
}

\begin{document}

\title{Mirror Symmetry and \\ Double Signature Change}

\author{
	Tevian Dray
	\footnote{Permanent address:
	Department of Mathematics,
	Oregon State University,
	Corvallis, OR  97331, USA, 
	\texttt{tevian@math.oregonstate.edu}.
	}
	\\[-2.5pt]
	\normalsize\textit{African Institute for Mathematical Sciences}
		\\[-2.5pt]
	\normalsize\textit{Muizenberg 7950, SOUTH AFRICA} \\[-2.5pt]
	\normalsize\url{tevian@aims.ac.za} \\[10pt]
	Charles Hellaby \\[-2.5pt]
	\normalsize\textit{Department of Mathematics and Applied Mathematics}
		\\[-2.5pt]
	\normalsize\textit{University of Cape Town} \\[-2.5pt]
	\normalsize\textit{Rondebosch 7701, SOUTH AFRICA} \\[-2.5pt]
	\normalsize\url{Charles.Hellaby@uct.ac.za}
}

\date{\normalsize \today}

\maketitle

\begin{abstract}
The \textit{black mirror} spacetime proposed by Tzanavaris, Boyle, and Turok~\cite{Boyle25} connects the two exterior regions of the extended Schwarzschild black hole directly to each other, with no intervening interior region.  Using techniques adapted from previous work on signature change, we reexamine the black mirror spacetime as a model of \textit{double signature change}, and investigate whether there is a surface layer at the horizon, that is, a distributional curvature singularity corresponding to an impulsive gravitational wave.  We confirm that the black mirror spacetime does not contain any such singularity, and compare our result with previous claims that the curvature components are analytic.  We also discuss the global structure of the black mirror spacetime, and examine what happens to worldlines and curves passing through.
\end{abstract}

\newpage

\section{Introduction}

Although the Schwarzschild solution was discovered in 1916~\cite{Schwarzschild16}, and was shown to be the unique spherically symmetric vacuum solution of Einstein's field equation in 1921~\cite{Jebsen21,Birkhoff23}, the physical interpretation of this solution has a long and tortured history.  Already in 1917, Droste~\cite{Droste17} gave the modern interpretation of $r$ as the radial coordinate, and in 1924 Eddington~\cite{Eddington24} showed implicitly that there is no physical singularity at the horizon, but it was only in the 1960s, after the work of Szekeres~\cite{Szekeres60} and Kruskal~\cite{Kruskal60}, that the structure of the horizon at $r=2m$ was fully understood.  Or so we thought...

The \textit{black mirror} spacetime proposed in 2024 by Tzanavaris, Boyle, and Turok~\cite{Boyle25} adds a new twist to this history.  Motivated by the role of the \textit{Euclidean} Schwarzschild geometry in the thermodynamic properties of black holes, in this spacetime the two \textit{exterior} regions of the maximally extended Kruskal spacetime are joined to each other across the horizon, without passing through an intermediate interior region.

What are the physical properties of the resulting joined horizon?  Although the spacetime is a vacuum solution everywhere else, it is not obvious how to analyze the horizon itself.  Coordinate-based approaches run the risk of hiding physical singularities behind coordinate singularities, especially if the metric itself is not well-behaved in the chosen coordinates.

Similar issues arise in the analysis of impulsive gravitational waves, where small coordinate shifts can result in gravitational shock waves.  The situation is even more extreme in the case of signature change, in which a Euclidean region is joined to a Lorentzian region, where even the underlying differentiable structure depends on the assumptions made.

We argue here that a careful analysis of the horizon structure in the black mirror spacetime suggests that there is indeed \textit{no} surface layer at the horizon, corresponding to an impulsive gravitational wave, despite initial indications otherwise.  As described below, our analysis is based on treating the horizon as a surface of \textit{double signature change}.

We summarize previous results in Section~\ref{background}, discussing the black mirror spacetime, impulsive gravitational waves, and (single) signature change.  We then present our analysis of the black mirror spacetime in terms of double signature change in Section~\ref{double}, analyze the global structure of the black mirror spacetime in Section~\ref{global} (with some details relegated to two appendices), and discuss our results in Section~\ref{discuss}, where we also point out several unresolved issues that could be topics for future work.

\section{Background}
\showlabel{background}

\subsection{Black Mirrors}
\showlabel{mirror}

The Schwarzschild metric in ingoing/outgoing Eddington--Finkelstein coordinates is
\begin{equation}
ds^2 = -\left(1-\frac{2m}{r}\right)dv_\pm^2 \pm 2\,dv_\pm\,dr + r^2d\Omega^2 ~,
\showlabel{EF}
\end{equation}
where $d\Omega^2$ denotes the 2-sphere metric, which smoothly extends the original Schwarzschild spacetime across the horizon at $r=2m$.  Since~\eqref{EF} is nondegenerate at $r=2m$, this extension demonstrates that $r=2m$ is merely a coordinate singularity in Schwarzschild coordinates, but not a curvature singularity.

Replacing $r$ by a new coordinate $\sigma$ defined as in~\cite{Boyle25} by
\begin{equation}
r = 2m \left( 1 + \frac{\sigma^2}{16m^2} \right)
\end{equation}
brings the line element to the form
\begin{equation}
ds^2 = -\frac{2m}{r} \left(\frac{\sigma}{4m}\right)^2 dv_\pm^2
	\pm \frac{\sigma}{2m}\,dv_\pm\,d\sigma + r^2d\Omega^2
\showlabel{black}
\end{equation}
with the horizon $r=2m$ corresponding to $\sigma=0$.  Since $r\ge 2m$ for all values of $\sigma$, allowing $\sigma$ to take on both positive and negative values leads to the interpretation in~\cite{Boyle25} that the two exterior regions of the maximally extended Schwarzschild spacetime have been joined at $\sigma=0$, with no interior region.

As noted in~\cite{Boyle25}, the metric~\eqref{black} is clearly a vacuum solution of Einstein's equation away from $\sigma=0$.  Since~\eqref{black} is smooth, these authors claim that this \textit{black mirror} spacetime contains no curvature singularities.  However, it is relevant to the discussion that follows that the black mirror metric~\eqref{black} is degenerate at the horizon $\sigma=0$, unlike the Eddington--Finkelstein metric~\eqref{EF}.  In particular, it is not obvious whether the connection and curvature at the horizon are well defined.

\subsection{Constructing the Black Mirror Spacetime}
\showlabel{constrmirror}

The metric \eqref{black} is intentionally double-valued in~$\sigma$, and Tzanavaris, Boyle, and Turok~\cite{Boyle25} propose that positive~$\sigma$ represents {\rI} of the Penrose diagram, while negative~$\sigma$ represents {\rIII}.  Their construction is intended to address some questions in black hole information theory.  In this paper, however, we are interested in several questions about the classical spacetime created by this construction.  Are the junctions well-defined?  What is the character of these junctions?  What is the resulting global structure?

For our investigation, it is convenient to work with the double-null version of \KrSz\ coordinates~\cite{Szekeres60,Kruskal60}, and use Penrose~\cite{Penr64,HawEll73,Wald84} coordinates for the diagrams.  The metric becomes
 \begin{align}
   ds^2 & = -\frac{32 m^3}{r} e^{-r/2m} \, dU \, dV + r^2 \, d\Omega^2 ,
 \showlabel{eq:KSm}
 \end{align}
where $r$ is defined implicitly by
 \begin{align}
   e^{r/2m} \left( \frac{r}{2m} - 1 \right) = - U V ~.
 \showlabel{eq:R2MUV}
 \end{align}
The corresponding (standard) Penrose diagram is shown in Figure~\ref{PenDg}.
The time coordinate used by \KrSz\ is given by
 \begin{align}
   T = \frac{1}{2} \big( U + V \big) ~.
 \end{align}
In these coordinates,  the spherical vacuum spacetime is manifestly well-behaved for all values of $r$ other than $0$, and the time ($T$) direction is upward throughout.

\goodbreak
The construction given in~\cite{Boyle25} (using slightly different labeling) proceeds as follows: (i)~{\rII} and {\rIV}, not including the horizons, are removed, (ii)~the upper horizon of {\rI} is joined to the lower horizon of {\rIII}, so that $(0,V)$ maps to $(0,-V)$, and (iii)~the lower horizon of {\rI} is joined to the upper horizon of {\rIII}, so that $(U,0)$ maps to $(-U,0)$.  In order to better understand what is happening at the first such junction, Figure~\ref{Flip1} shows the two regions glued together along the horizons that have been identified; the second junction is essentially the same as this one.  After the second pair of horizons is joined, we get the ``pita pocket'' of Figure~\ref{Pita1}.

\begin{center}


\psset{unit=30mm,xunit=30mm,yunit=30mm}
\pspicture(-2.05,-1.2)(2.05,1.2)

\psset{linewidth=0.3mm,linecolor=black}

\FPset{\sx}{1}			
\FPset{\sy}{1}
\rput(0,0){\RII}
\FPset{\sx}{-1}
\FPset{\sy}{-1}
\FPset{\ag}{90}
\rput(0,0.38){II}

\FPset{\sx}{1}			
\FPset{\sy}{-1}
\rput(0,0){\RII}
\FPset{\sx}{-1}
\FPset{\sy}{-1}
\FPset{\ag}{270}
\rput(0,-0.38){IV}

\FPset{\sx}{1}			
\FPset{\sy}{1}
\FPset{\ag}{0}
\rput(0,0){\RI}
\rput(0,0){\imzp}
\FPset{\sx}{1}
\FPset{\sy}{1}
\FPset{\ag}{0}
\rput(0.75,0){I}

\FPset{\sx}{-1}			
\FPset{\sy}{1}
\FPset{\ag}{0}
\rput(0,0){\RI}
\rput(0,0){\imzp}
\FPset{\sx}{1}
\FPset{\sy}{1}
\FPset{\ag}{180}
\rput(-0.75,0){III}

\FPset{\su}{1}
\FPset{\sv}{1}
\rput(0,-0.04){\UVarrows}

\endpspicture
\\
\pb{13cm}{
\refstepcounter{figure}
\showlabel{PenDg}
{\small
Figure~\arabic{figure}.~~
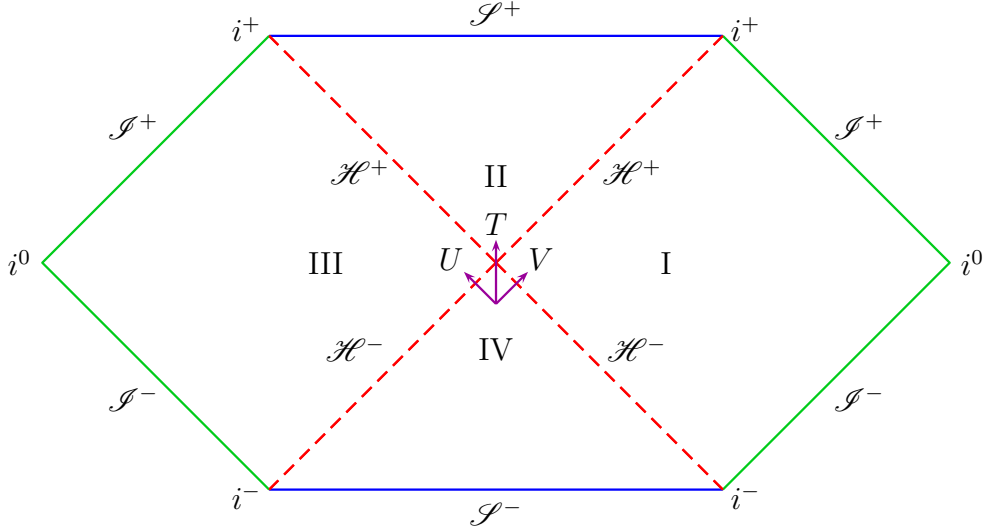
The standard Penrose diagram for the maximally extended Schwarzschild spacetime.  The directions of the null \KrSz\ coordinates $(U,V)$ are shown; the angular coordinates $(\theta,\phi)$ are suppressed.  The upward arrow represents the global time orientation given by the \KrSz\ time coordinate $T$  (although $T$ is not everywhere vertical).
}
}


\psset{unit=30mm,xunit=30mm,yunit=30mm}
\pspicture(-1.15,-1.15)(2.2,2.2)

\psset{linewidth=0.3mm,linecolor=black}

\FPset{\sx}{1}			
\FPset{\sy}{1}
\FPset{\ag}{0}
\rput{\ag}(0,0){\RI}
\rput{\ag}(0,0){\imzp}
\FPset{\su}{1}
\FPset{\sv}{1}
\rput(0.8,0){\UVarrows}
\FPset{\sx}{1}
\FPset{\sy}{1}
\FPset{\ag}{0}

\FPset{\sx}{-1}			
\FPset{\sy}{-1}
\FPset{\ag}{270}
\rput{\ag}(0,0){\RI}
\rput{\ag}(0,0){\imzp}
\FPset{\su}{-1}
\FPset{\sv}{1}
\rput(0,0.8){\UVarrows}
\FPset{\sx}{1}
\FPset{\sy}{-1}
\FPset{\ag}{90}

\multido{\i=1+2}{5}{		
  \FPeval{\mp}{0.1*\i}		
  \FPeval{\smjax}{\mp*1.414}
  \FPdiv{\smnax}{\smjax}{2.5}
  \FPeval{\lcr}{1*\mp}
  \FPeval{\lcg}{1 - \mp}
  \definecolor{mlc}{rgb}{\lcr,0.5,\lcg}
  \rput{-45}(0,0){
    \psellipticarc[linecolor=mlc]{->}(0,0)(\smjax,\smnax){180}{360}
  }
}

\endpspicture
\\
\pb{13cm}{
\refstepcounter{figure}
{
\small Figure~\arabic{figure}\showlabel{Flip1}.~~
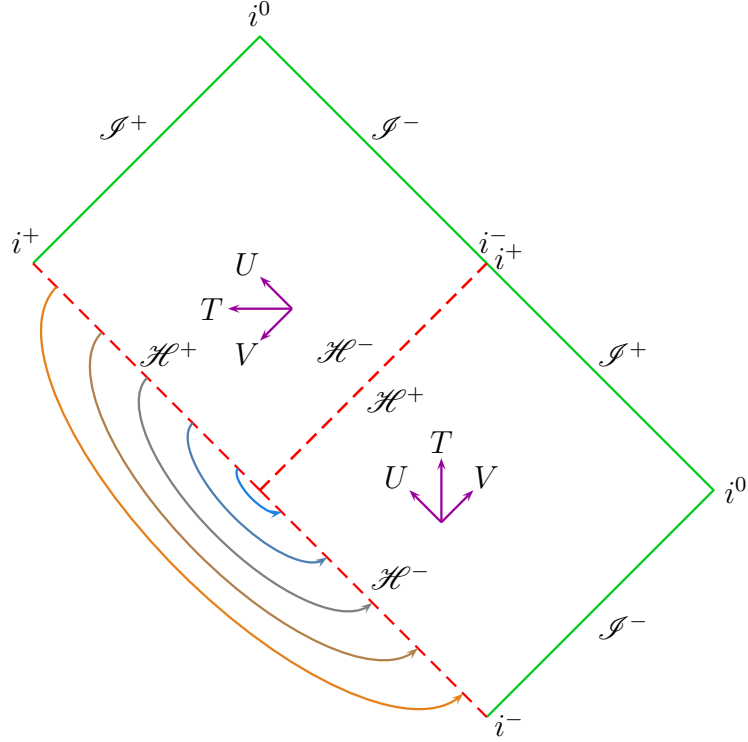
Starting from the Penrose diagram of Figure~\ref{PenDg}, after removing Regions II and IV, {\rIII} has been flipped about the $U$ axis and joined to region {\rI} along their horizons at $U = 0$, as in~\cite{Boyle25}.  The mapping used to join the remaining horizons at $V=0$ is also shown.  Notice that the $T$ direction in the flipped {\rIII} is now (approximately) horizontal; see Section~\ref{cvcont}.
}
}


 \psset{unit=38mm, xunit=38mm, yunit=38mm}
 \pspicture(-0.05, -1.1)(2.05, 1)
 \FPset{\RD}{57.29577950} 
 \FPset{\Al}{20}
 \FPset{\Bt}{10}
 \FPeval{\Alph}{clip(-\Al + 90)}                                                       
 \psset{Alpha=\Alph,Beta=\Bt}
 \FPset{\rttw}{1.414}       
 \FPset{\rtthtw}{0.8660}    
 \FPset{\pifr}{0.7854}      
 \FPset{\pith}{1.047}       
 \FPset{\cosC}{0.6837}
 \FPset{\sinC}{0.7297}
 \parametricplotThreeD[linecolor=red,linestyle=dashed](0,\rttw){                                        
   \rtthtw\space \pith\space \rtthtw\space mul COS mul \cosC\space mul \sinC\space 2 div sub t mul
   \rtthtw\space \pith\space \rtthtw\space mul COS mul \sinC\space mul \cosC\space 2 div add t mul
   t \rtthtw\space mul \pith\space \rtthtw\space mul SIN mul
 }
 \parametricplotThreeD[linecolor=red,linestyle=dashed](0,\rttw){                                        
   \rtthtw\space \pith\space \rtthtw\space mul COS mul \cosC\space mul \sinC\space 2 div sub t mul
   \rtthtw\space \pith\space \rtthtw\space mul COS mul \sinC\space mul \cosC\space 2 div add t mul
   t \rtthtw\space mul \pith\space \rtthtw\space mul SIN mul neg
 }
 \parametricplotThreeD[linecolor=Green](0,\pifr){                                                       
   2 \rtthtw\space mul 4 3 div t mul \rtthtw\space mul COS mul \cosC\space mul \sinC\space sub t COS t SIN add div
   2 \rtthtw\space mul 4 3 div t mul \rtthtw\space mul COS mul \sinC\space mul \cosC\space add t COS t SIN add div
   2 \rtthtw\space mul 4 3 div t mul \rtthtw\space mul SIN mul t COS t SIN add div
 }
 \parametricplotThreeD[linecolor=Green](-\pifr,0){                                                       
   2 \rtthtw\space mul 4 3 div t mul \rtthtw\space mul COS mul \cosC\space mul \sinC\space sub t COS t SIN sub div
   2 \rtthtw\space mul 4 3 div t mul \rtthtw\space mul COS mul \sinC\space mul \cosC\space add t COS t SIN sub div
   2 \rtthtw\space mul 4 3 div t mul \rtthtw\space mul SIN mul t COS t SIN sub div
 }
 \parametricplotThreeD[linecolor=Green](0,\pifr){                                                       
   2 \rtthtw\space mul 4 3 div t mul \rtthtw\space mul COS mul \cosC\space mul \sinC\space sub t COS t SIN add div neg
   2 \rtthtw\space mul 4 3 div t mul \rtthtw\space mul COS mul \sinC\space mul \cosC\space add t COS t SIN add div
   2 \rtthtw\space mul 4 3 div t mul \rtthtw\space mul SIN mul t COS t SIN add div
 }
 \parametricplotThreeD[linecolor=Green](-\pifr,0){                                                       
   2 \rtthtw\space mul 4 3 div t mul \rtthtw\space mul COS mul \cosC\space mul \sinC\space sub t COS t SIN sub div neg
   2 \rtthtw\space mul 4 3 div t mul \rtthtw\space mul COS mul \sinC\space mul \cosC\space add t COS t SIN sub div
   2 \rtthtw\space mul 4 3 div t mul \rtthtw\space mul SIN mul t COS t SIN sub div
 }
 \pstThreeDLine[linecolor=purple]{->}(-0.4545,1.75,-0.1)(-0.4545,1.85,0)
 \pstThreeDLine[linecolor=purple]{->}(-0.4545,1.75,-0.1)(-0.4545,1.65,0)
 \pstThreeDLine[linecolor=purple]{->}(-0.4545,1.75,-0.1)(-0.4545,1.75,0.07)
 \pstThreeDLine[linecolor=purple]{->}(0.4545,1.75,0.1)(0.4545,1.85,0)
 \pstThreeDLine[linecolor=purple]{->}(0.4545,1.75,0.1)(0.4545,1.65,0)
 \pstThreeDLine[linecolor=purple]{->}(0.4545,1.75,0.1)(0.4545,1.75,-0.07)
 \pstThreeDPut[pOrigin=t](-0.5,1.5,0.55){$\Scri^+$}
 \pstThreeDPut[pOrigin=t](0.5,1.5,0.45){$\Scri^-$}
 \pstThreeDPut[pOrigin=t](-0.5,1.5,-0.5){$\Scri^-$}
 \pstThreeDPut[pOrigin=t](0.5,1.5,-0.35){$\Scri^+$}
 \pstThreeDPut[pOrigin=t](-0.2,0.6,0.53){$\Scrh^+$}
 \pstThreeDPut[pOrigin=t](0.2,0.6,0.47){$\Scrh^-$}
 \pstThreeDPut[pOrigin=t](-0.2,0.61,-0.46){$\Scrh^-$}
 \pstThreeDPut[pOrigin=t](0.2,0.61,-0.25){$\Scrh^+$}
 \pstThreeDPut[pOrigin=t](-0.03,1.13,1.05){$i^+$}
 \pstThreeDPut[pOrigin=t](0.03,1,1.05){$i^-$}
 \pstThreeDPut[pOrigin=t](-0.03,1.1,-0.95){$i^-$}
 \pstThreeDPut[pOrigin=t](0.03,0.97,-0.95){$i^+$}
 \endpspicture
\\
\pb{13cm}{
\refstepcounter{figure}
{
\small Figure~\arabic{figure}\showlabel{Pita1}.~~
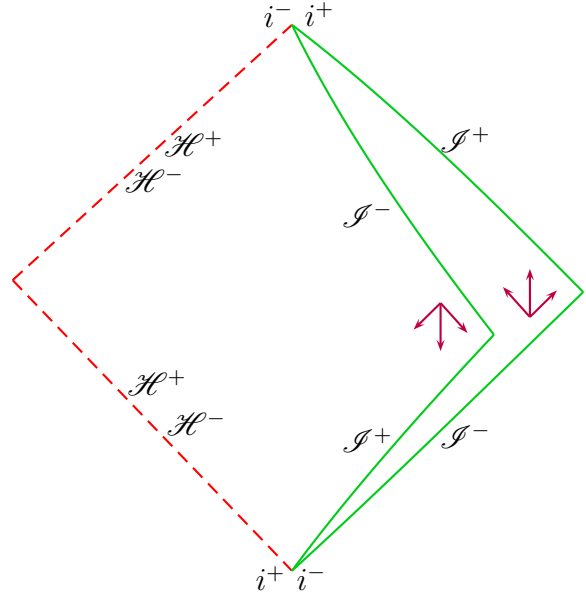
The forward pita-pocket diagram.  {\rIII} has been flipped a second time about the $V$ axis, and joined to the bottom of {\rI} according to the mapping shown in Figure~\ref{Flip1}.  Notice that $T$ flows up in the back sheet and down in the front sheet.  The future horizon of each region has been joined to the past horizon of the other.
}
}

\end{center}

In Figure~\ref{PenDg}, it is evident that time $T$ flows upward throughout the diagram.  However, we note both that the standard mapping between Schwarzschild and \KrSz\ coordinates has the Schwarzschild time coordinate $t$ going down in {\rIII}, and that, once Regions II and IV have been cut out, the natural linkage of the future directions in Regions I and III of the \KrSz\ spacetime has been severed.~%
\footnote{Since there are still infinitely many curves through $(U,V) = (0,0)$, going between {\rI} and {\rIII}, along which one could require the continuity of orientation, this may not be entirely true.  However, we have treated the two joined regions as independent.}

In the context of our classical model, it is natural for a black horizon, into which (test) matter is falling, to be joined to a white horizon, which is ejecting matter.  This ``\textit{forward}'' scenario is depicted in Figures~\ref{Flip1}--\ref{Pita1}.  There is a consistent time direction, derived from the \KrSz\ time coordinate~$T$, flowing up one side of the pita, down the other, and back round again.

By contrast,~\cite{Boyle25} label the upper horizon in {\rIII} as ``WH'' (white hole) and the lower horizon as ``BH'' (black hole), with the local time orientation chosen to match the Schwarzschild time coordinate $t$.  This ``\textit{reverse}'' arrangement, in which the two black horizons are joined to each other, and similarly the two white horizons, is also possible.

An equivalent identification is shown in Figure~\ref{AltMap}, again using the time orientation given by $T$.  Causally, (test) matter from both regions is disappearing into a common black horizon, where worldlines just seem to end, and similarly, a shared white horizon is feeding (test) matter, apparently out of nowhere, into two distinct regions.  This arrangement is depicted in the reverse  pita-pocket diagram, Figure~\ref{Pita2}.

\begin{center}


\psset{unit=30mm,xunit=30mm,yunit=30mm}
\pspicture(-2.05,-1.1)(2.05,1.15)

\psset{linewidth=0.3mm,linecolor=black}

\FPset{\sx}{1}           
\FPset{\sy}{1}
\FPset{\ag}{0}
\rput(0,0){\RI}
\FPset{\su}{1}
\FPset{\sv}{1}
\rput(0.8,0){\UVarrows}
\FPset{\sx}{1}
\FPset{\sy}{1}
\FPset{\ag}{0}

\rput[lb](1.01,1.01){$v = 0$}

\FPset{\sx}{-1}           
\FPset{\sy}{1}
\FPset{\ag}{0}
\rput(0,0){\RI}
\FPset{\su}{1}
\FPset{\sv}{1}
\rput(-0.8,0){\UVarrows}
\FPset{\sx}{1}
\FPset{\sy}{1}
\FPset{\ag}{180}

\psset{linewidth=0.3mm,linecolor=black}

\multido{\i=1+2}{5}{		
  \FPeval{\mp}{0.1*\i}		
  \FPeval{\ar}{\mp*2.3}
  \FPeval{\as}{\ar/2}
  \FPeval{\lcr}{1*\mp}
  \FPeval{\lcg}{1 - \mp}
  \definecolor{mlc}{rgb}{\lcr,0.5,\lcg}
  \psellipticarc[linecolor=mlc]{->}(0,0)(\ar,\as){45}{135}
  \psellipticarc[linecolor=mlc]{->}(0,0)(\ar,\as){225}{315}
}

\endpspicture
\\
\pb{13cm}{
\refstepcounter{figure}
{
\small Figure~\arabic{figure}\showlabel{AltMap}.~~
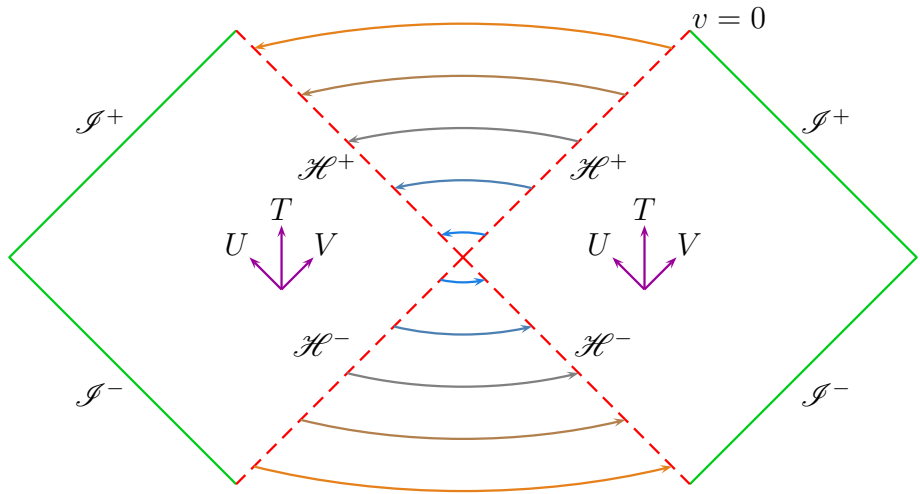
An alternative identification of the future and past horizons of \rI\ and \rIII.
}
}


 \psset{unit=38mm, xunit=38mm, yunit=38mm}
 \pspicture(-0.05, -1.1)(2.05, 1)
 \FPset{\RD}{57.29577950} 
 \FPset{\Al}{20}
 \FPset{\Bt}{10}
 \FPeval{\Alph}{clip(-\Al + 90)}                                                       
 \psset{Alpha=\Alph,Beta=\Bt}
 \FPset{\rttw}{1.414}       
 \FPset{\rtthtw}{0.8660}    
 \FPset{\pifr}{0.7854}      
 \FPset{\pith}{1.047}       
 \FPset{\cosC}{0.6837}
 \FPset{\sinC}{0.7297}
 \parametricplotThreeD[linecolor=red,linestyle=dashed](0,\rttw){                                        
   \rtthtw\space \pith\space \rtthtw\space mul COS mul \cosC\space mul \sinC\space 2 div sub t mul
   \rtthtw\space \pith\space \rtthtw\space mul COS mul \sinC\space mul \cosC\space 2 div add t mul
   t \rtthtw\space mul \pith\space \rtthtw\space mul SIN mul
 }
 \parametricplotThreeD[linecolor=red,linestyle=dashed](0,\rttw){                                        
   \rtthtw\space \pith\space \rtthtw\space mul COS mul \cosC\space mul \sinC\space 2 div sub t mul
   \rtthtw\space \pith\space \rtthtw\space mul COS mul \sinC\space mul \cosC\space 2 div add t mul
   t \rtthtw\space mul \pith\space \rtthtw\space mul SIN mul neg
 }
 \parametricplotThreeD[linecolor=Green](0,\pifr){                                                       
   2 \rtthtw\space mul 4 3 div t mul \rtthtw\space mul COS mul \cosC\space mul \sinC\space sub t COS t SIN add div
   2 \rtthtw\space mul 4 3 div t mul \rtthtw\space mul COS mul \sinC\space mul \cosC\space add t COS t SIN add div
   2 \rtthtw\space mul 4 3 div t mul \rtthtw\space mul SIN mul t COS t SIN add div
 }
 \parametricplotThreeD[linecolor=Green](-\pifr,0){                                                       
   2 \rtthtw\space mul 4 3 div t mul \rtthtw\space mul COS mul \cosC\space mul \sinC\space sub t COS t SIN sub div
   2 \rtthtw\space mul 4 3 div t mul \rtthtw\space mul COS mul \sinC\space mul \cosC\space add t COS t SIN sub div
   2 \rtthtw\space mul 4 3 div t mul \rtthtw\space mul SIN mul t COS t SIN sub div
 }
 \parametricplotThreeD[linecolor=Green](0,\pifr){                                                       
   2 \rtthtw\space mul 4 3 div t mul \rtthtw\space mul COS mul \cosC\space mul \sinC\space sub t COS t SIN add div neg
   2 \rtthtw\space mul 4 3 div t mul \rtthtw\space mul COS mul \sinC\space mul \cosC\space add t COS t SIN add div
   2 \rtthtw\space mul 4 3 div t mul \rtthtw\space mul SIN mul t COS t SIN add div
 }
 \parametricplotThreeD[linecolor=Green](-\pifr,0){                                                       
   2 \rtthtw\space mul 4 3 div t mul \rtthtw\space mul COS mul \cosC\space mul \sinC\space sub t COS t SIN sub div neg
   2 \rtthtw\space mul 4 3 div t mul \rtthtw\space mul COS mul \sinC\space mul \cosC\space add t COS t SIN sub div
   2 \rtthtw\space mul 4 3 div t mul \rtthtw\space mul SIN mul t COS t SIN sub div
 }
 \pstThreeDLine[linecolor=purple]{->}(-0.4545,1.75,-0.1)(-0.4545,1.85,0)
 \pstThreeDLine[linecolor=purple]{->}(-0.4545,1.75,-0.1)(-0.4545,1.65,0)
 \pstThreeDLine[linecolor=purple]{->}(-0.4545,1.75,-0.1)(-0.4545,1.75,0.07)
 \pstThreeDLine[linecolor=purple]{->}(0.4545,1.75,-0.1)(0.4545,1.85,0)
 \pstThreeDLine[linecolor=purple]{->}(0.4545,1.75,-0.1)(0.4545,1.65,0)
 \pstThreeDLine[linecolor=purple]{->}(0.4545,1.75,-0.1)(0.4545,1.75,0.07)
 \pstThreeDPut[pOrigin=t](-0.5,1.5,0.55){$\Scri^+$}
 \pstThreeDPut[pOrigin=t](0.5,1.5,0.45){$\Scri^+$}
 \pstThreeDPut[pOrigin=t](-0.5,1.5,-0.5){$\Scri^-$}
 \pstThreeDPut[pOrigin=t](0.5,1.5,-0.35){$\Scri^-$}
 \pstThreeDPut[pOrigin=t](-0.2,0.6,0.53){$\Scrh^+$}
 \pstThreeDPut[pOrigin=t](0.2,0.6,0.47){$\Scrh^+$}
 \pstThreeDPut[pOrigin=t](-0.2,0.61,-0.46){$\Scrh^-$}
 \pstThreeDPut[pOrigin=t](0.2,0.61,-0.25){$\Scrh^-$}
 \pstThreeDPut[pOrigin=t](-0.03,1.13,1.05){$i^+$}
 \pstThreeDPut[pOrigin=t](0.03,1,1.05){$i^+$}
 \pstThreeDPut[pOrigin=t](-0.03,1.1,-0.95){$i^-$}
 \pstThreeDPut[pOrigin=t](0.03,0.97,-0.95){$i^-$}
 \endpspicture
\\
\pb{13cm}{
\refstepcounter{figure}
{
\small Figure~\arabic{figure}\showlabel{Pita2}.~~
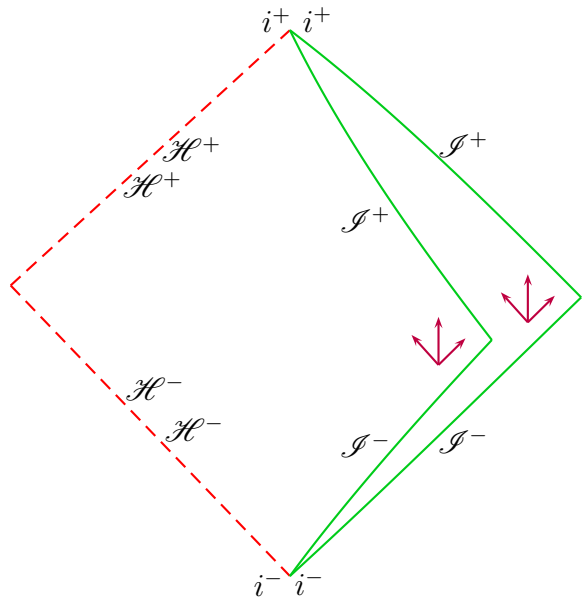
The reverse pita-pocket diagram.  In this case, {\rIII} is flipped just once  about the vertical ($T$) axis, and joined with {\rI} along both the upper and lower horizons.  Notice that $T$ flows upward in both the back and front sheets---the two future horizons are merged, and the two past horizons are merged.
}
}

\end{center}

These mappings of one horizon to another cause some curious effects.  The $U$ and $V$ directions in Figure~\ref{Flip1} indicate that what was a timelike direction in {\rI} extends into a spacelike direction in {\rIII}, while a spacelike direction in {\rI} continues as a timelike direction {\rIII}.  These effects will be discussed further in Section~\ref{cvcont}.

\subsection{Impulsive Gravitational Waves}
\showlabel{wave}

As noted at the end of Section~\ref{mirror}, it is not obvious whether the connection and curvature of the black mirror spacetime are well defined.  Similar issues arise with impulsive gravitational waves.

Penrose~\cite{Penrose72} showed how small shifts in the coordinates when crossing a null surface connecting two flat regions of spacetime could lead to distributional curvature, that is, to an impulsive gravitational wave propagating along the surface.  This theme recurs in many contexts, such as the work by~\cite{ASexl}, as well as a series of papers coauthored by one of us~\cite{DtH85a,DtH85b,DtH86}.

All of these impulsive waves can be described as the result of joining two spacetimes along a null surface, so it is natural to ask what the appropriate junction conditions are.  In the spacelike case, the appropriate boundary conditions are well known (see e.g.~\cite{Darm27,Isra66,Lich55}), and numerous authors have discussed the generalization to null boundaries (see e.g.~\cite{Dautcourt64,Penrose72,Taub73}).  We follow here, in Section~\ref{junction}, the unified presentation in Clarke and Dray~\cite{Clarke}.

\subsection{Signature Change}
\showlabel{sig}

In the early 1990s,~\cite{Tucker91,Tucker93} proposed a model of particle production in the context of quantum field theory in curved spacetime that was due to the presence of a Euclidean region, and~\cite{ElSuCoHe92,Elli92} independently proposed a mathematically similar model describing early universe cosmology.  Lengthy discussion followed~\cite{HelDra94,HelDra95,DraHel96,Patch}, leading eventually to the realization that there are two inequivalent models of signature change.

In \textit{continuous} signature change, the metric becomes degenerate at the junction, containing a term such as $-t\,dt^2$ that changes sign continuously.  In \textit{discontinuous} signature change, one avoids the degeneracy of the metric by working with a smooth orthonormal frame, but allows the metric to be discontinuous, so that the metric contains a term such as $\mp dt^2$.  In both cases, the intrinsic metric at the junction is assumed to be continuous and nondegenerate.

As was shown in~\cite{DrElHeMa97}, these two approaches correspond to inequivalent differentiable structures on the underlying manifold.  It is therefore reasonable to investigate both approaches, so long as one is clear about which case one is considering.  However, as discussed at the end of~\cite{Patch} and also noted in~\cite{Mars93}, the construction in~\cite{Clarke} shows that, even in the context of signature change, the smooth differentiable structures on either side of the junction extend uniquely to a $C^1$ differentiable structure that includes the boundary, namely the one that underlies \textit{discontinuous} signature change.  Thus, adopting the junction conditions from~\cite{Clarke} implicitly assumes that one has also chosen to work with discontinuous signature change.  Put differently, when joining two manifolds-with-boundary using \textit{continuous} signature change, the resulting differentiable structure can not agree with the original structures on both sides.

\subsection{Junction Conditions}
\showlabel{junction}

\subsubsection*{Unified Approach}

As noted above, we follow the unified treatment of junction conditions given in~\cite{Clarke}.  Let $\Sigma$ be a surface dividing a spacetime $(M,g_{ab})$ into two smooth manifolds-with-boundary $M^\pm$.  Defining the step function $\Theta$ to be $1$ on $M^+$ and $0$ on $M^-$, the gradient $\delta_a$ of $\Theta$ is a vector-valued distribution, which can be expanded as
\begin{equation}
\delta_a = \delta n_a
\showlabel{ddef}
\end{equation}
for any choice of normal vector $n^a$ to $\Sigma$; the Dirac delta ``function'' $\delta$ depends on the choice of $n^a$, but $\delta_a$ does not.

A tensor field $Q$ is \textit{regularly discontinuous} at $\Sigma$ if $Q$ is continuous on the interior of $M^\pm$, and if the one-sided limits $Q^\pm\big\vert_\Sigma$ of $Q$ to $\Sigma$ in $M^\pm$ exist.  In this case, the \textit{discontinuity} of $Q$ is the tensor on $\Sigma$ defined by
\begin{equation}
[Q]_\Sigma = Q^+\big\vert_\Sigma - Q^-\big\vert_\Sigma . 
\end{equation}
Note that $Q$ need not be defined on $\Sigma$.

In the null case, we have
\begin{equation}
n^a n_a = 0
\end{equation}
and we can construct a ``dual'' null vector $l^a$ satisfying~%
\footnote{Unlike the spacelike case, there is no canonical normalization of $n^a$ (and hence $l^a$) in the null case.}
\begin{align}
l^a l_a &= 0 ,\\
l^a n_a &= -1
\end{align}
on $\Sigma$.  The spacelike 2-surface $S$ to which both $n^a$ and $l^a$ are orthogonal has metric
\begin{equation}
q_{ab} = g_{ab} + 2l_{(a}n_{b)}
\end{equation}
where the round brackets denote symmetrization as usual.  We can now define several second fundamental forms associated with $S$, namely
\begin{align}
\chi_{ab} &= \nabla_c n_d q^c{}_a q^d{}_b \showlabel{chieq}\\
\psi_{ab} &= \nabla_c l_d q^c{}_a q^d{}_b \showlabel{psieq}\\
\eta_a &= \nabla_c l_d q^c{}_a n^d = -\nabla_c n_d q^c{}_a l^d
\showlabel{etaeq}
\end{align}
We will refer to $\chi_{ab}$ as \textit{internal}, to $\psi_{ab}$ as \textit{external}, and to $\eta_a$ as \textit{normal}.

Writing the stress-energy tensor as
\begin{equation}
T_{ab} = (1-\theta) T^-_{ab} + \theta T^+_{ab} + \delta\tau_{ab}
\end{equation}
and introducing
\begin{equation}
\omega^a = n^c\nabla_c n^a
\showlabel{omegaeq}
\end{equation}
to measure the extent to which $n^a$ is geodesic on $\Sigma$, it is shown in~\cite{Clarke} that
\begin{equation}
\tau_{ab}
  = -[\tr\psi]_\Sigma\,n_an_b -2[\eta_{(a}]_\Sigma\,n_{b)} -[\omega]_\Sigma\,q_{ab}
\showlabel{CDtau}
\end{equation}
where $\omega=-l_a\omega^a$.  Thus, the conditions for the absence of a surface layer at $\Sigma$ are 
\begin{equation}
[\tr\psi]_\Sigma = [\eta_a]_\Sigma = [\omega]_\Sigma = 0 .
\showlabel{CDjoin}
\end{equation}

\subsubsection*{Variational Approach}

It is not obvious that the derivation given above using the unified approach in~\cite{Clarke} is valid in the presence of signature change.  Since the metric is now discontinuous, there is in principle a distributional term in the connection, which has been ignored.

A resolution of this issue for a spacelike boundary was proposed in~\cite{JMP} using a variational principle.  As in~\cite{Clarke}, one assumes that two manifolds-with-boundary are joined along the identified boundary, then assumes that the theory is derived from the Einstein--Hilbert action with boundary terms.  In the spacelike case, one does indeed recover the Darmois junction conditions for the lack of a surface layer.  In retrospect, it is not surprising that this variational approach contains no distributional connection term, but otherwise agrees with the unified approach.  So it seems reasonable to anticipate a similar conclusion in the null case.

However, the identified boundaries can not be null on both sides (and isometric) unless either there is no signature change or, as considered here, double signature change.  In this case, the derivation of~\eqref{CDtau} and~\eqref{CDjoin} is even more problematic than in the case of a spacelike surface.  We therefore take the variational approach as fundamental, without attempting to compare the resulting junction conditions with those in~\cite{Clarke} in general, although we will nonetheless do so for the specific example of the black mirror.

So we assume that we are given two smooth manifolds-with-boundary $M^\pm$ with smooth, nondegenerate metrics $g^\pm$, whose diffeomorphic boundaries $\Sigma^\pm$ have been identified.  We will further assume that this identification results in a smooth manifold $M$~%
\footnote{As discussed in~\cite{Clarke}, it is sufficient for $M^\pm$ to be $C^3$, for $g^\pm$ to be $C^2$, and for $M$ to be merely $C^1$ at $\Sigma$.}
divided by a smooth hypersurface $\Sigma\cong\Sigma^\pm$ into disjoint manifolds that can be identified with the interiors of $M^\pm$.  We assume that the pullbacks of $g^\pm$ to $\Sigma$ agree, but we allow the pullbacks to be degenerate, that is, $\Sigma$ could be a null surface.  The only such case we consider, referred to henceforth simply as $\Sigma$ being null, is where both of $M^\pm$ are Lorentzian.

As discussed in~\cite{JMP}, if $\Sigma$ is spacelike, an orthonormal basis of 1-forms in $\Sigma$ can be smoothly extended to an orthonormal basis of 1-forms on $M$ by appending a smooth choice of 1-form normal to $\Sigma$.  If $\Sigma$ is null, a similar process can be used to construct a smooth, double-null basis of 1-forms on $M$, as follows.

Let $n$ be a 1-form normal to $\Sigma$, so that $n(X)=0$ for any tangent vector $X\in T\Sigma$.  Choose a null 1-form $l\in T^*\Sigma$ satisfying $|g(n,l)|=|g^{ab}n_al_b|=1$.  (The sign ambiguity is necessary in the presence of signature change.)  Assuming $M$ is 4-dimensional, there is a 2-dimensional submanifold $S$ of $\Sigma$ such that both $n$ and $l$ are normal to $S$.  Choose an orthonormal basis $\{m,k\}$ of $T^*S$.  Then the desired double-null basis of 1-forms is obtained by extending $\{n,l,m,k\}$ off of $\Sigma$.  As an example, such a basis in Minkowski space would be $\{du,dv,dx,dy\}$, with $u=(t-z)/\sqrt2$, $v=(t+z)/\sqrt2$, and $\Sigma=\{u=0\}$.

In order for integration on $\Sigma$ to be well defined, we will assume that our double-null basis has been constructed so as to be compatible with the orientation of $M$, and then use the induced orientation on $\Sigma$.  In other words, we assume that the volume element of $M$ is
\begin{equation}
\omega = n \wedge l \wedge m \wedge k
\end{equation}
and that the induced orientation on $\Sigma$ is such that $l\wedge m\wedge k$ is ``positive''.  Returning to the example of Minkowski space, these assumptions would imply that
$\omega=du\wedge dv\wedge dx\wedge dy=dt\wedge dx\wedge dy\wedge dz$, thus determining the relative sign in the relation
\begin{equation}
\int_\Sigma dv\wedge dx\wedge dy = +\int_\Sigma dv\,dx\,dy
\end{equation}
that defines the integration of differential forms on $\Sigma$.

The Einstein--Hilbert Lagrangian density on a manifold-with-boundary is given by
\begin{equation}
\mathcal{L}
  = g_{ac}\Omega^c{}_b\wedge{*}(e^a\wedge e^b)
    - d\left( g_{ac} \omega^c{}_b\wedge{*}(e^q\wedge e^b) \right)
\end{equation}
where $e^a$ are the orthonormal basis 1-forms, $\omega^a{}_b$ are the connection 1-forms, and $\Omega^a{}_b$ are the curvature 2-forms, given by
\begin{equation}
\Omega^a{}_b = d\omega^a{}_b + \omega^a{}_c \wedge \omega^c{}_b .
\end{equation}
We will assume that the connection is metric-compatible, that is, that
\begin{equation}
dg_{ab} = \omega^m{}_a g_{mb} + \omega^m{}_b g_{ma}  .
\showlabel{metcomp}
\end{equation}
Restricting to double-null bases, the metric is constant, that is, $dg_{ab}=0$, as is also the case for orthonormal bases.  It is well known (see e.g.~\cite{JMP}) that varying the action $\int_M\mathcal{L}$ with respect to the connection leads to the condition that the connection also be torsion-free, that is, that
\begin{equation}
de^a + \omega^a{}_b\wedge e^b = 0 ,
\showlabel{torfree}
\end{equation}
so that the connection is the Levi-Civita connection.  Similarly, it is well known (see e.g.~\cite{JMP}) that varying the action with respect to the frame yields Einstein's vacuum equations away from the boundary.

Of interest here is the variation of the surface term with respect to the frame.  As shown in~\cite{JMP}, we obtain
\begin{equation}
-\int_\Sigma \delta_e \left( g_{ac} \omega^c{}_b\wedge{*}(e^q\wedge e^b) \right)
= \int_\Sigma \rho_d \wedge \delta e^d
\showlabel{evar}
\end{equation}
where
\begin{equation}
\rho_d = g_{ac} \omega^c{}_b \wedge {*}(e^a\wedge e^b\wedge e^mg_{md}) .
\end{equation}
Thus, the condition that the joined manifold $M$ satisfy the equations of motion derived from the Einstein--Hilbert action (with boundary) is that the surface terms in $M^\pm$ must cancel, that is, that $\rho^\pm_d$ must agree on $\Sigma$.  It is worth emphasizing that the integration in~\eqref{evar} has the effect of pulling back the integrand to $\Sigma$.  Thus, the junction conditions for the absence of a surface layer at a null boundary are therefore precisely that the \textit{pullbacks} of $\rho^\pm_d$ to $\Sigma$ must agree.  We write this condition on the (lack of) discontinuity in (the pullback of) $\rho_d$ as
\begin{equation}
[\rho_d] = 0  .
\showlabel{drho}
\end{equation}

\section{Double Signature Change}
\showlabel{double}

Starting with the Schwarzschild line element in double-null \KrSz\ coordinates \eqref{eq:KSm}, \eqref{eq:R2MUV}, we consider the ``forward'' identification, taking {\rI} as $M^-$ and {\rIII} as $M^+$, and we join them along the horizon at $U=0$ by identifying $V$ in \rI\ with $-V$ in \rIII, which has the effect of changing the sign of the first term in~\eqref{eq:KSm} as well as the overall sign of~\eqref{eq:R2MUV}.  Explicitly, we now have
\begin{equation}
ds^2 = - \epsilon\, \frac{32m^3}r e^{-r/2m}dU\,dV + r^2 d\Omega^2 ~,
\showlabel{KSpm}
\end{equation}
where 
\begin{equation}
UV = - \epsilon\, e^{r/2m}\left(\frac{r}{2m}-1\right) ~,
\showlabel{UVpm}
\end{equation}
while
\begin{equation}
\mbox{$\epsilon = +1$ in $M^-$ ~~and~~ $\epsilon = -1$ in $M^+$} ~,
\showlabel{Hdef}
\end{equation}
so that $\epsilon=+1$ corresponds to the standard double-null form~\eqref{eq:KSm}.

We refer to this sign change as \textit{double signature change}, since two components of the metric in an orthonormal frame change sign at the junction.
We emphasize that the (new) coordinates $(U,V,\theta,\phi)$ are smooth by construction, corresponding to the unique differentiable structure constructed in~\cite{Clarke}.  The metric~\eqref{KSpm} is of course discontinuous.

Using the relations
\begin{align}
\rU &= - \epsilon\, \frac{4m^2}{r}e^{-r/2m} V ,\showlabel{rU}\\
\rV &= - \epsilon\, \frac{4m^2}{r}e^{-r/2m} U  \showlabel{rV}
\end{align}
it is straightforward to show that the only independent Christoffel symbols which are discontinuous at $\Sigma$ are
\begin{align}
\Gamma^U{}_{UU} &= \epsilon\, \frac{2m}{r}\left(1+\frac{2m}{r}\right)e^{-r/2m}V ,
\showlabel{Gam1}\\
\Gamma^\theta{}_{\theta U} = \Gamma^\phi{}_{\phi U} 
  &= - \epsilon\, \frac{4m^2}{r^2}e^{-r/2m}V .
\showlabel{Gam2}
\end{align}

\subsection{Is there a Surface Layer?}
\showlabel{surface}

\subsubsection*{Direct Computation}

We first attempt a direct, formal computation of the curvature starting from~\eqref{KSpm}, which we rewrite in terms of the Heaviside (step) function $\Theta=\Theta(U)$ as
\begin{equation}
ds^2 = (2\Theta-1)\frac{32m^3}r e^{-r/2m}dU\,dV + r^2 d\Omega^2 ,
\showlabel{KST}
\end{equation}
where~\eqref{UVpm} now becomes~%
\footnote{For later convenience, we have moved the step function to the left in~\eqref{UVT}, which is equivalent to noting that \textit{as distributions} $(2\Theta-1)^2=1$, that is, $\Theta(1-\Theta)$ is the zero distribution.  Although $\epsilon = 1 - 2\Theta$ away from $U = 0$, $\epsilon$ is not a distribution.}
\begin{equation}
UV (2\Theta-1) = e^{r/2m}\left(\frac{r}{2m}-1\right) .
\showlabel{UVT}
\end{equation}
Differentiating implicitly now yields~\eqref{rU} and~\eqref{rV} in the form
\begin{align}
\rU &= (2\Theta-1+2U\delta)\frac{4m^2}{r}e^{-r/2m} V ,\showlabel{rUT}\\
\rV &= (2\Theta-1)\frac{4m^2}{r}e^{-r/2m} U  \showlabel{rVT}
\end{align}
where $\delta=\delta(U)=\Theta'(U)$ is the Dirac delta distribution.  Since $U\delta$ is the zero distribution, we can eliminate that term from~\eqref{rUT}, resulting in the \textit{weak derivative} of $r$ with respect to $U$.

Working in coordinates $(U,V,\theta,\phi)$ and using~\eqref{rUT} and~\eqref{rVT}, we can now calculate the Christoffel symbols, discovering that only $\Gamma^U{}_{UU}$ is truly distributional, that is, contains~$\delta$.  Although one can not normally compute the curvature tensor of a distributional connection, in this case the only independent curvature component $R^a{}_{bcd}$ that fails to be continuous is
\begin{equation}
R^U{}_{UUV}
  = \frac{4m}{r^4}e^{-r/2m}\bigl(8m^3(2\Theta-1) + (r+2m)r^2U\delta\bigr) .
\showlabel{Rie}
\end{equation}
Eliminating the zero distribution $U\delta$ as before, we conclude that the Riemann tensor is in fact merely discontinuous, rather than distributional.  Furthermore, raising an index in~\eqref{Rie} when computing the Ricci tensor will remove the discontinuity (due to the discontinuity of the metric); the components $R_{ab}$ of the Ricci tensor are continuous.  At this point, it should come as no surprise that the Ricci tensor vanishes, as, of course, does the Einstein tensor.

This formal computation at the very least suggests that there is no surface layer, that is, that the black mirror spacetime is a vacuum solution of Einstein's equation everywhere, including at the horizons.

\subsubsection*{Variational Approach}

As in Section~\ref{double}, we use the line element \eqref{KSpm} with \eqref{UVpm} and \eqref{Hdef}, in which $V$ has been replaced with $-V$ in \rIII.  Introducing
\begin{equation}
f(r) = 4\sqrt{\frac{m^3}{r}}e^{-r/4m} \showlabel{fDef}
\end{equation}
a smooth double-null frame is given by
\begin{align}
n = e^U &= f\,dU ~, \showlabel{nDefKS} \\
l = e^V &= f\,dV ~, \showlabel{lDefKS} \\
m = e^\theta &= r\,d\theta ~, \showlabel{eThDefKS} \\
k = e^\phi &= r\,\sin\theta\,d\phi ~, \showlabel{ePhDefKS}
\end{align}
and it is straightforward to check that the nonzero components of the (inverse) metric in this non-coordinate basis are
\begin{equation}
g^{UV} = - \epsilon = g^{VU} , \qquad g^{\theta\theta} = 1 = g^{\phi\phi}
\end{equation}
as desired.  A smooth orientation (volume element) on $M$ is given by
\begin{equation}
\omega
  = e^U \wedge e^V \wedge e^\theta \wedge e^\phi
  = f^2 r^2 \sin\theta \, dU \wedge dV \wedge d\theta \wedge d\phi
\showlabel{4vol}
\end{equation}
which induces the orientation on $\Sigma$ such that
\begin{equation}
\sigma
  = e^V \wedge e^\theta \wedge e^\phi
  = f r^2 \sin\theta \, dV \wedge d\theta \wedge d\phi
\showlabel{3vol}
\end{equation}
is positive.

We emphasize that the orientation $\sigma$ in~\eqref{3vol} is induced by~\eqref{4vol} in both of $M^\pm$, and also that this orientation of $M^+$ (and hence also the orientation it induces on $\Sigma$) nonetheless differs from the original \KrSz\ orientation in \rIII, due to the replacement of (the original) $V$ with $-V$ in that region.

It is now straightforward to work out the Hodge dual acting on our basis of 1-forms, yielding
\begin{align}
{*}e^U &= - \epsilon\, e^V \wedge e^\theta \wedge e^\phi ,\\
{*}e^V &= + \epsilon\, e^U \wedge e^\theta \wedge e^\phi ,\\
{*}e^\theta &= +e^U \wedge e^V \wedge e^\phi ,\\
{*}e^\phi &= -e^U \wedge e^V \wedge e^\theta ,
\end{align}
which can be used to compute the Hodge dual acting on 3-forms, since $** = +1$ on 1-forms in four dimensions and Lorentzian signature.

Since both factors in~\eqref{drho} are antisymmetric in the indices $a$ and $b$, direct computation yields
\begin{align}
\rho_U &= 2\bigl( \omega^V{}_\theta \wedge e^\phi
	- \omega^V{}_\phi \wedge e^\theta
	+ \omega^\theta{}_\phi \wedge e^V \bigr) ,\showlabel{rho1}\\
\rho_V &= 2\bigl( -\omega^U{}_\theta \wedge e^\phi
	+ \omega^U{}_\phi \wedge e^\theta
	- \omega^\theta{}_\phi \wedge e^U \bigr) ,\\
\rho_\theta &= 2\bigl( \epsilon\, \omega^V{}_V \wedge e^\phi
	+ \omega^V{}_\phi \wedge e^U
	- \omega^U{}_\phi \wedge e^V \bigr) ,\\
\rho_\phi &= 2\bigl( - \epsilon\, \omega^V{}_V \wedge e^\theta
	- \omega^V{}_\theta \wedge e^U
	+ \omega^U{}_\theta \wedge e^V \bigr) .
\showlabel{rho2}
\end{align}

Turning to the connection, direct computation from~\eqref{metcomp} and~\eqref{torfree} yields
\begin{align}
\omega^U{}_U
  &= \frac{1}{f^2} \fU e^U - \frac{1}{f^2} \fV e^V
   = -\omega^V{}_V ,\showlabel{om1}\\
\omega^U{}_\theta &= \epsilon\, \frac{1}{rf} \rV e^\theta ,\showlabel{omUth}\\
\omega^U{}_\phi &= \epsilon\, \frac{1}{rf} \rV e^\phi ,\\
\omega^V{}_\theta &= \epsilon\, \frac{1}{rf} \rU e^\theta ,\\
\omega^V{}_\phi &= \epsilon\, \frac{1}{rf} \rU e^\phi ,\showlabel{om2}\\
\omega^\theta{}_\phi &= -\frac{\cot\theta}{r} e^\phi .
\end{align}

Without further computation, we see from~\eqref{rU} and~\eqref{rV} that the each of the partial derivatives in~\eqref{om1}--\eqref{om2} is discontinuous, adding an extra factor of $\epsilon = \mp 1$ (on $M^\pm$) in each case, which either cancels the existing factor of $\epsilon$ in~\eqref{omUth}--\eqref{om2}, or, upon insertion into~\eqref{rho1}--\eqref{rho2}, precisely cancels the factors of $\epsilon$ appearing there.  Thus, $\rho_d$ is continuous!  This is more than we need to satisfy~\eqref{drho}, as we haven't yet pulled $\rho_d$ back to $\Sigma$ (which is essentially just setting $e^U$ to $0$).

Thus, there is no distributional term in the variation of the Einstein--Hilbert action at the horizon, and hence no surface layer.

\subsubsection*{Unified Approach}

In the presence of signature change, the construction of $\delta_a$ in~\eqref{ddef} implicitly assumes that the 1-form $n_a$ is continuous.
We choose $n_a$ and $l_a$ as in~\eqref{nDefKS} and~\eqref{lDefKS}, so that $n_U=l_V=f$ (with $f$ given by~\eqref{fDef}).  Since $f$ is constant on the 2-surface $S$, the fundamental forms \eqref{chieq}--\eqref{etaeq} become
\begin{align}
\chi_{ij} &= -f\Gamma^U{}_{ij} ,\\
\psi_{ij} &= -f\Gamma^V{}_{ij} ,\\
\eta_i &= - \epsilon \frac1f\Gamma^U{}_{Ui} = 0
\end{align}
where $i$ and $j$ run over the coordinates $(\theta,\phi)$ on $S$.  Comparison with~\eqref{Gam1} and~\eqref{Gam2} shows that each of these fundamental forms is continuous at $\Sigma$.  Similarly, since $n^V=\pm1/f$, \eqref{omegaeq} becomes
\begin{equation}
\omega^a
= \frac1f\frac{\partial}{\partial V}\left(\frac1f\right) \delta^a{}_V
  + \frac1{f^2} \Gamma^a{}_{VV}
\end{equation}
which is also continuous at $\Sigma$, since the derivative~\eqref{rV} is zero there.

We conclude that each term in~\eqref{CDtau} vanishes, so that the conditions~\eqref{CDjoin} are satisfied and there is no surface layer at $\Sigma$.

\subsection{Comparison}
\showlabel{compare}

Three different computations were summarized in Section~\ref{surface}, using direct computation, a variational approach, and the unified junction conditions in~\cite{Clarke}, respectively.  Only the variational approach rests on a solid mathematical footing, although it is reassuring that the other two approaches nonetheless appear to yield the same result.  When working with distributions, it is often the case that naive direct calculations give the correct answer so long as the result, as here, does not contain unmanageable terms such as $\delta^2$.

The applicability of the null junction conditions given in~\cite{Clarke} is more problematic.  The derivation of~\eqref{CDtau} as given in~\cite{Clarke} assumes that
\begin{equation}
[g_{ab,c}] = n_c \gamma_{ab} ,
\end{equation}
allowing the discontinuities in the Christoffel symbols to be expressed as
\begin{equation}
2[\Gamma^a{}_{bc}] = n_c \gamma^a{}_b + n_b \gamma^a{}_c - n^a \gamma_{bc}
\showlabel{CDgam}
\end{equation}
resulting in
\begin{equation}
2[\psi_{ab}] = \gamma_{cd} q^c{}_a q^d{}_b .
\end{equation}
Thus,
\begin{equation}
[\tr\psi] = \gamma^\theta{}_\theta + \gamma^\phi{}_\phi
\end{equation}
and it would appear from~\eqref{Gam2} that
\begin{equation}
n_U \gamma^\theta{}_\theta
  = -[\Gamma^\theta{}_{\theta U}]
  \ne 0
\end{equation}
so that $[\tr\psi]\ne0$, resulting in a surface layer.~%
\footnote{We originally did precisely this computation, incorrectly concluding that the black mirror spacetime contained an impulsive gravitational wave.}

However, the discontinuities in $n^a$ and $l^a$ must be taken into account in~\eqref{etaeq} and~\eqref{omegaeq}.  If we instead compute
\begin{equation}
n^V \gamma_{\theta\theta}
  = 2 [\Gamma^V{}_{\theta\theta}]
  = 0
\end{equation}
we obtain a different conclusion, namely that there is no surface layer, in agreement with our other computations.  In a nutshell, care must be taken when raising indices in the presence of signature change.  It is largely for this reason that we have not been able to compare the results of the variational approach with the null junction conditions given in~\cite{Clarke} in the general case.

\section{Global Structure}
\showlabel{global}

\subsection{Double Joining}
\showlabel{join2}

As in~\cite{Boyle25}, we can repeat this construction along the past horizon of \rI, joining it to the future horizon of \rIII.  The curvature computation in Section~\ref{double} does not need to be repeated, as this scenario differs from that one by simply swapping $U$ and $V$.  Thus, there is no surface layer at $V=0$.  The resulting ``pita pocket'' geometry of the black mirror spacetime is shown in Figure~\ref{Pita1}.

\subsection{Time Orientation}
\showlabel{time}

In Section~\ref{double}, we implicitly assumed that \rIII\ lies to the future of \rI, thus respecting the global time orientation inherited from the maximally extended Schwarzschild spacetime.  However, since these regions share only a single sphere (where the horizons cross), we could have reversed the time orientation of \rIII\ instead.  In this scenario, the horizon becomes a future boundary, which does not appear to be physically reasonable.  Since the Schwarzschild geometry is time symmetric, an equivalent scenario could also be obtained by gluing together the two future horizons.  Similarly, in the double joining considered in Section~\ref{join2}, there would be both a future boundary along one identified horizon, and a past boundary along the other.

\subsection{Curve Continuation}
\showlabel{cvcont}

At a Lorentzian-to-Euclidean signature change, timelike or null worldlines on one side are forced to become spacelike on the other.  Indeed, by joining a Lorentzian spacetime to a Euclidean space it is implied that the regular matter content turns into something entirely different.  Although the manifolds joined here are both Lorentzian, the way they are joined actually creates a double signature change, in the sense that a curve passing from $M^-$ to $M^+$ that is smooth relative to the $(U, V)$ coordinates, may change from timelike to spacelike along some directions, and spacelike to timelike along others.

The mismatch between the timelike directions of $M^-$ and $M^+$ at these two junctions presents us with a difficulty.  Given a family of timelike worldlines arriving at $\Sigma$, how should they be continued through the junction?  Should they just continue ``straight ahead'', as Newton's laws would seem to require, or should they stay timelike, to preserve a causal arrow of time?  As we'll see, both options have unphysical aspects.

Consider a curve with affine parameter $\alpha$ and tangent vector $w^a = \p x^a / \p \alpha$, arriving at $\Sigma$ from one side.  Since there are metric sign changes at the junction, the spacelike, null or timelike character of a curve could also change there, in which case the squared magnitude $w^a w_a$ also changes.  It is therefore too restrictive to consider only ``unit speed'' parameterizations of (non-null) curves, that is, timelike curves parameterized by proper time (whose tangent vector is the 4-velocity) and spacelike curves parameterized by arclength.  We therefore work throughout this section with the components of the ``4-momentum'' $P^a = \mu w^a$.  We impose the following conditions:
\begin{itemize}
\item[]   (i)~~The tangential components $P^a e^i_a$ and $P^a \ell_a$ along $\Sigma$ are preserved (see \eqref{nDefKS}--\eqref{ePhDefKS});
\item[]   (ii)~~The transverse component $P^a n_a$ stays the same;
\item[]   (iii)~~If the direction (as determined by the contravariant components of $P^a$) is toward $\Sigma^-$ on the $M^-$ side, then it is away from $\Sigma^+$ on the $M^+$ side , or vice-versa.  (Equivalently, the curve's affine parameter $\alpha$ is monotonic through $\Sigma$.)
\end{itemize}
In the absence of signature change, it is common and often simpler to replace (ii) with
\begin{itemize}
\item[]   (ii$'$)~~The squared magnitude $P^a P_a$ stays the same,
\end{itemize}
because both versions of (ii) are equivalent, and (i)--(iii) ensure that, in coordinates based on the junction surface, all components of $P^a$ and $w^a$ are kept the same.  When the signature changes however, as we'll see, difficulties arise.  Thus, we further insist that:
\begin{itemize}
\item[]   (iv)~~The curve continuation requirements must connect every distinct curve arriving at or leaving one side with a unique curve on the other side.
\end{itemize}

The case of single signature change at a spacelike 3-surface in a black hole metric was considered in~\cite{HeSuEl97}.  It was found there that only the equivalent of (ii) allows all possible paths crossing $\Sigma$ to be propagated through.

In the case of the horizon matchings contemplated here, timelike and spacelike paths are possible on each side.  We {\it could\/} implement a version of (ii$'$), and keep timelike paths timelike.  But, as shown below, we can only do this by putting sharp kinks in the paths (see Fig.~\ref{GeodJoinLK} and case (c) in the following, and compare with Fig.~\ref{Flip1}); and one might reasonably ask why the signature changes in the metric do not affect affect the character of geodesics or test matter or other paths.  For paths traversing this double signature change in the joined-up metric, it is entirely likely that $P^a P_a$ changes sign as well as magnitude at $\Sigma$.

We will call the curve-continuation option using (ii) the double signature change (DSC) option, and the one using (ii$'$) the Lorentzian kink (LK) option.  

\subsubsection{Double Signature Change Continuation}
\showlabel{cvcalc}

We consider the junction that matches the two parts of $U = 0$, and has surface coordinates $\xi^i = (V,\theta,\phi)$.  By \eqref{nDefKS}, \eqref{lDefKS} and \eqref{fDef}, the radial basis forms are
\begin{align}
   n_a = \left( f, 0, 0, 0 \right) ~,~~~~~~ \ell_a = \left( 0, f, 0, 0 \right) ~,~~~~~~ f = 4 \sqrt{\frac{m^3}{r}} e^{-r/4m} ~.
 \showlabel{eq:nlf}
\end{align}
In this subsection, all quantities are evaluated on the junction surface, where
\begin{align}
   U_\Sigma = 0 ~,~~~~~~ r_\Sigma = 2 m ~,~~~~~~ f_\Sigma = 2 m \sqrt{\frac{2}{e}} ~.
\end{align}
For a particle path or curve $\gamma$ with 4-momentum $P^a = \mu w^a$ arriving at $\Sigma$, we have
 \begin{align}
   P^a e^i_a = \big( P^\theta, P^\phi \big) ~,~~~~~~ & P^a \ell_a = f P^V
 \showlabel{Pcomps}
 \end{align}
which we ideally want to preserve across $\Sigma$ in the global coordinates.

We use the sign factor $\epsilon$ of \eqref{Hdef}, such that $\epsilon = +1$ in $M^-$ and $-1$ in $M^+$, writing the metric components as
\begin{align}
   g_{ab} =
      \begin{pmatrix}
      0 & - \epsilon f^2 & 0 & 0 \\
      - \epsilon f^2 & 0 & 0 & 0 \\
      0 & 0 & r^2 & 0 \\
      0 & 0 & 0 & r^2 \sin^2(\theta)
      \end{pmatrix}
      ~,
\end{align}
and we use $s = \pm 1, 0$ for the spacelike, timelike or null character of the path.  Thus, we find
 \begin{align}
   P^a P_a = \mu^2 s & = - 2 f^2 \, \epsilon \, P^U \, P^V
      + r^2 \Big( \big( P^\theta \big)^2 + \sin^2\theta \big( P^\phi \big)^2 \Big) 
 \showlabel{eq:PSq} \\
   P^a n_a & = f \, P^U
 \showlabel{eq:Pn} \\
   P^a \ell_a & = f \, P^V
 \showlabel{eq:Pl}
 \end{align}
For $P^a \ell_a$ to be constant, eq \eqref{eq:Pl} ensures $P^V$ stays the same through $\Sigma$, so we will continue to assume this.  Note that in the context of this matching, $U < 0$ in region {\rI} and $U > 0$ in region {\rIII}, so condition (iii) requires $P^U > 0$ on both sides, and eq \eqref{eq:Pn} confirms $P^a n_a$ is continuous.  We consider a sequence of cases to build up the picture.

(a) Null and radial (incoming in $M^-$, outgoing in $M^+$)
 \begin{align}
   P^a & = \big( P^U, 0, 0, 0 \big) ~~~~~~\to \\
   \mu^2 s & = 0 ~,~~~~ P^a \ell_a = 0 ~.
 \end{align}
Only $P^a n_a$ relates the new and old $P^U$.  This is the single direction that stays null through $\Sigma$, see Figure~\ref{NullCones}.

(b) Timelike or spacelike and radial
 \begin{align}
   P^a & = \big( P^U, P^V, 0, 0 \big) ~~~~~~\to \\
   \mu^2 s & = - 2 f^2 \, \epsilon \, P^U \, P^V ~.
 \end{align}
Clearly timelike becomes spacelike and conversely.

(c) Null and not radial.  (Obviously, purely tangential is not possible for a light ray at $r = 2 m$, and there is a limit on its angular speed.)
 \begin{align}
   P^a & = \big( P^U, P^V, P^\theta, 0 \big) ~~~~~~\to
 \showlabel{eq:cP} \\
   \mu^2 s & = - 2 f^2 \, \epsilon \, P^U \, P^V + r^2 \big( P^\theta \big)^2
 \showlabel{eq:cPP}
 \end{align}
If the curve were to stay null, $\mu^2 s = 0$, eq \eqref{eq:cPP} would make $P^U$ change its sign, which violates requirement (iii).  But (ii) with \eqref{eq:Pn} also says that $P^U$ stays the same, so we must conclude that a path that is null on one side becomes spacelike on the other, except for the special case (a).  Further, it is evident that preserving $P^a P_a$ or $w^a w_a$ is not possible, so we discard (ii$'$) as a continuation condition under double signature change.

(d) Timelike or spacelike and not radial \\
The equations are the same as \eqref{eq:cP}--\eqref{eq:cPP}, but $\mu^2 s \neq 0$ on one side at least.  Obviously timelike always switches to spacelike, but spacelike may connect to timelike, null or even spacelike if the angular component is large enough, that is if
 \begin{align}
   4 m^2 \big( P^\theta \big)^2 > \frac{16 m^2}{e} \, \big| P^U \, P^V \big| ~.
 \showlabel{eq:SplToSpl}
 \end{align}
We emphasise that these results apply whichever way the path is extended---forward into $M^+$ or backward into $M^-$.  

These results are illustrated in Figure~\ref{NullCones}, which shows how such horizon junctions cause the signature to change in opposite senses in two distinct (sets of) directions.  This is why the junction is a ``double signature change''.

\begin{center}


 \psset{unit=35mm, xunit=35mm, yunit=35mm}
 \pspicture(-1.5,-1.3)(2.57,1.25)
 \FPset{\RD}{57.29577950} 
 \FPset{\Al}{-30}
 \FPset{\Bt}{10} 
 \FPeval{\Alph}{clip(-\Al + 90)}                                                       
 \psset{Alpha=\Alph,Beta=\Bt}

 \FPset{\ncr}{1}
 \FPset{\ql}{0.2}
 \FPset{\pl}{1.6}

 \parametricplotThreeD[linecolor=blue,xPlotpoints=300](0,360){                                        
   t cos \ncr\space mul
   t sin \ncr\space mul
   \ncr\space neg
 }
 \FPeval{\aco}{72/\RD} 
 \FPeval{\xco}{\ncr*cos(\aco)}
 \FPeval{\yco}{\ncr*sin(\aco)}
 \pstThreeDLine[linecolor=blue](\xco,\yco,-\ncr)(0,0,0)
 \FPeval{\aco}{230/\RD} 
 \FPeval{\xco}{\ncr*cos(\aco)}
 \FPeval{\yco}{\ncr*sin(\aco)}
 \pstThreeDLine[linecolor=blue](\xco,\yco,-\ncr)(0,0,0)
 \pstThreeDLine[linecolor=cyan,linestyle=dashed](0,0,0)(0,-\ncr,-\ncr)
 \pstThreeDPut[pOrigin=t](0,1,-1.28){PNC in {\rI}}

 \parametricplotThreeD[linecolor=Green,xPlotpoints=300](0,360){                                        
   t cos \ncr\space mul
   \ncr\space neg
   t sin \ncr\space mul
 }
 \FPeval{\aco}{69/\RD} 
 \FPeval{\xco}{\ncr*cos(\aco)}
 \FPeval{\zco}{\ncr*sin(\aco)}
 \pstThreeDLine[linecolor=Green](\xco,-\ncr,\zco)(0,0,0)
 \FPeval{\aco}{317/\RD} 
 \FPeval{\xco}{\ncr*cos(\aco)}
 \FPeval{\zco}{\ncr*sin(\aco)}
 \pstThreeDLine[linecolor=Green](\xco,-\ncr,\zco)(0,0,0)
 \pstThreeDLine[linecolor=cyan,linestyle=dashed](0,0,0)(0,-\ncr,-\ncr)
 \pstThreeDPut[pOrigin=b](0,-1,1.03){FNC in {\rIII}}

 \pstThreeDLine[linecolor=cyan](\ncr,-\ncr,-\ncr)(\ncr,\ncr,\ncr)(-\ncr,\ncr,\ncr)(-\ncr,-\ncr,-\ncr)(\ncr,-\ncr,-\ncr)     
 \pstThreeDPut[pOrigin=lb](0,\ncr,\ncr){Null junction surface $\Sigma$}

 \pstThreeDDot[linecolor=red](0,0,0)
 \pstThreeDPut[pOrigin=b](0,0,0.05){$X$}

 \pstThreeDLine[linecolor=magenta,arrowsize=0.06]{->}(\pl,\ql,0)(-\pl,-\ql,0)
 \pstThreeDPut[pOrigin=l](\pl,\ql,0){~S-like goes to S-like}

 \pstThreeDLine[linecolor=magenta,arrowsize=0.06]{->}(\ncr,\ncr,0)(-\ncr,-\ncr,0)
 \pstThreeDPut[pOrigin=l](\ncr,\ncr,0){~S-like goes to N}

 \pstThreeDLine[linecolor=magenta,arrowsize=0.06]{->}(\ql,\pl,0)(-\ql,-\pl,0)
 \pstThreeDPut[pOrigin=l](\ql,\pl,0){~S-like goes to T-like}


 \endpspicture
\\
\pb{13cm}{
\refstepcounter{figure}
{
\small Figure~\arabic{figure}\showlabel{NullCones}.~~
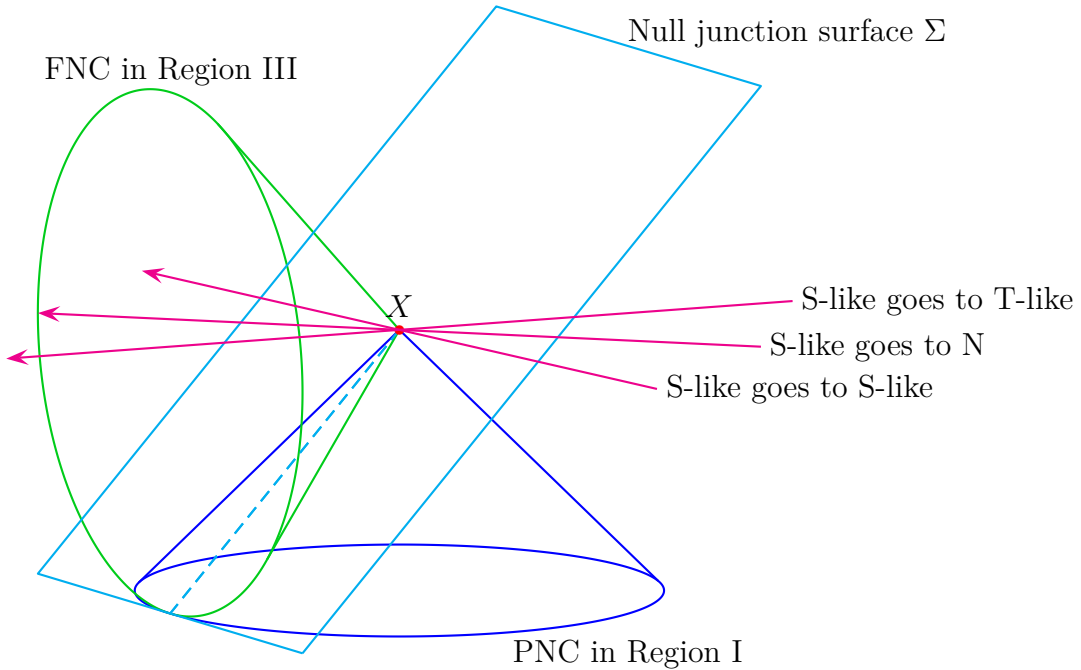
The two halves of the null cone of a point $X$ on $\Sigma$; they touch the null junction surface along the cyan dashed line.  ``FNC'' and ``PNC'' are the future and past null cones.  This is an elaboration of part of Figure~\ref{Flip1}, with a dimension added.  The three magenta paths all pass through $X$, going from {\rI} to {\rIII}, and the ones drawn here are at constant ``height'' in the diagram.  The diagram makes it obvious why spacelike paths can continue through a double signature change into timelike, null or spacelike paths.  Upward paths in and on the blue past null cone similarly all become spacelike.
}
}

\end{center}

\subsubsection{Lorentzian Kink Continuation}
\showlabel{LKcvcalc}

If we do not like the consequence that a timelike particle becomes tachyonic after encountering the ``black mirror'', then we could insist on timelike paths mapping to timelike paths, etc., but then there is a sharp kink in the path, as if $\Sigma$ delivers a massive impulse, see Figure~\ref{GeodJoinLK}.~%
\footnote{One may ask what could cause such a sudden change in direction?  Is there something on $\Sigma$ imparting a powerful impulse?  If there is some kind of light cannon it would have to point against the positive light flow direction on the horizon $U = 0$ to effect the correct momentum change.}
For an LK continuation, it is simplest to specify 
 \begin{align}
   \big( P^U, P^V, P^\theta, P^\phi) ~~\to~~ \big( P^U, -P^V, P^\theta, P^\phi) ~,
 \showlabel{LKcont}
 \end{align}
in other words $P^V$ flips its sign at $\Sigma$ in the global coordinates (and keeps its sign relative to the original coordinates in region {\rIII}).~%
\footnote{This may perhaps be the ``mirror'' effect implied in \cite{Boyle25}.}  
Since all curves retain their character, there is just one case.~%
\footnote{In principle there are many possible alternatives to \eqref{LKcont} --- timelike-to-timelike continuation rules consistent with (ii$'$) --- this is just the mathematically simplest.}

\subsubsection{Closed Timelike Geodesics?}
\showlabel{closedtimelike}

Each of the DSC and LK options allows us to connect geodesics across the two junctions, but because of the jump in either the components of the magnitude of the ``4-momentum'', they are piecewise geodesics.

In the double signature change curve-matching procedure, a radially infalling timelike geodesic goes into a spacelike geodesic and extends to $i^0$.  In the Lorentzian kink scenario, timelike radial geodesics continue as timelike geodesics.  These two alternatives are shown schematically in Figures~\ref{GeodJoinDSC}--\ref{GeodJoinLK} (paths not calculated).

\begin{center}


\noindent
\psset{unit=20mm,xunit=20mm,yunit=20mm}
\pspicture(-2.05,-1.1)(2.05,1.05)

\psset{linewidth=0.3mm,linecolor=black}

\FPset{\sx}{1}           
\FPset{\sy}{1}
\FPset{\ag}{0}
\rput{\ag}(0,0){\RI}
\rput{\ag}(0,0){\TLcurve}

\FPset{\sx}{-1}           
\FPset{\sy}{1}
\FPset{\ag}{0}
\rput{\ag}(0,0){\RI}
\rput{\ag}(0,0){\SLcurve}

\endpspicture
\hspace*{5mm}
\psset{unit=20mm,xunit=20mm,yunit=20mm}
\pspicture(-1.05,-1.05)(2.05,2.05)

\psset{linewidth=0.3mm,linecolor=black}

\FPset{\sx}{1}           
\FPset{\sy}{1}
\FPset{\ag}{0}
\rput{\ag}(0,0){\RI}
\rput{\ag}(0,0){\TLcurve}

\FPset{\sx}{-1}           
\FPset{\sy}{-1}
\FPset{\ag}{270}
\rput{\ag}(0,0){\RI}
\rput{\ag}(0,0){\SLcurve}

\endpspicture
\\
\pb{13cm}{
\refstepcounter{figure}
{
\small Figure~\arabic{figure}\showlabel{GeodJoinDSC}.~~
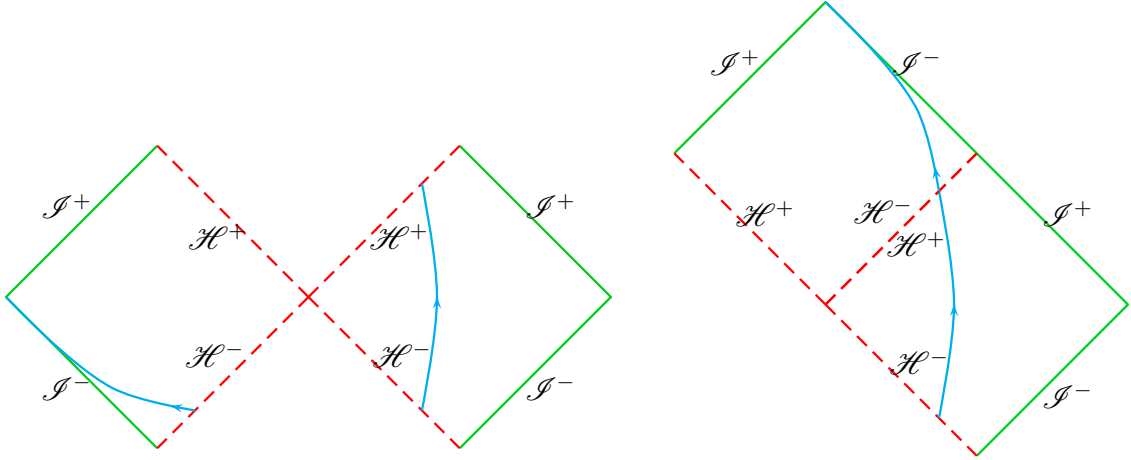
A timelike radial geodesic in {\rI} (cyan) continued ``straight through'' $\Sigma$ becomes a spacelike geodesic in {\rIII}.  This is the DSC case.  On the left, the geodesic is shown passing through Regions I and III of the Penrose diagram, and on the right {\rIII} has been flipped about the $U$ axis, as in Figure~\ref{Flip1}, to show the geodesic smoothly continued through $\Sigma$.
}
}


\noindent
\psset{unit=20mm,xunit=20mm,yunit=20mm}
\pspicture(-2.05,-1.1)(2.05,1.05)

\psset{linewidth=0.3mm,linecolor=black}

\FPset{\sx}{1}           
\FPset{\sy}{1}
\FPset{\ag}{0}
\rput(0,0){\RI}
\rput{\ag}(0,0){\TLcurve}

\FPset{\sx}{-1}           
\FPset{\sy}{1}
\FPset{\ag}{0}
\rput(0,0){\RI}
\rput{\ag}(0,0){\TLcurve}

\endpspicture
\hspace*{5mm}
\psset{unit=20mm,xunit=20mm,yunit=20mm}
\pspicture(-1.05,-1.05)(2.05,2.05)

\psset{linewidth=0.3mm,linecolor=black}

\FPset{\sx}{1}           
\FPset{\sy}{1}
\FPset{\ag}{0}
\rput{\ag}(0,0){\RI}
\rput{\ag}(0,0){\TLcurve}

\FPset{\sx}{-1}           
\FPset{\sy}{-1}
\FPset{\ag}{270}
\rput{\ag}(0,0){\RI}
\rput{\ag}(0,0){\TLcurve}

\endpspicture
\\
\pb{13cm}{
\refstepcounter{figure}
{
\small Figure~\arabic{figure}\showlabel{GeodJoinLK}.~~
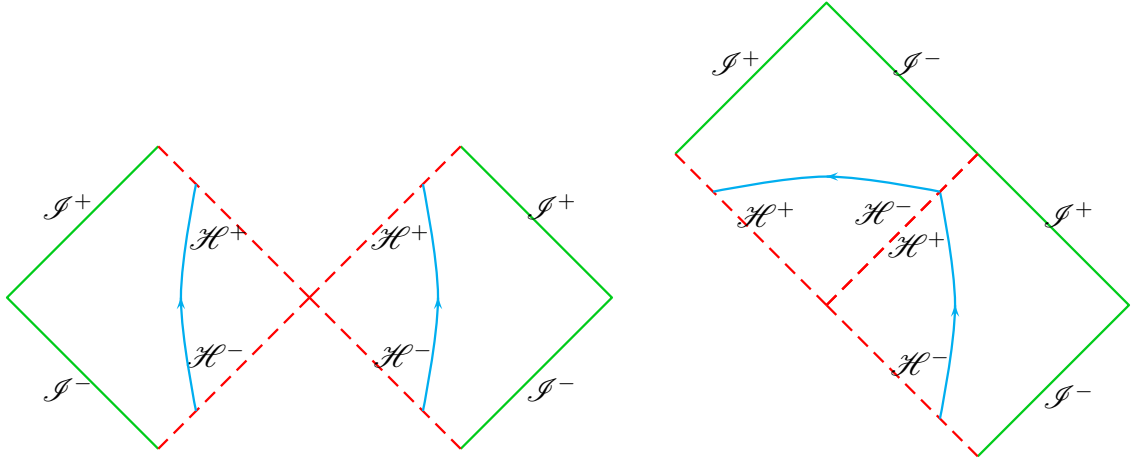
If one insists a timelike radial geodesic in {\rI} is continued through $\Sigma$ as another timelike geodesic in {\rIII}, the LK case, there has to be a kink at $\Sigma$, i.e.\ it suffers a violent acceleration there.  Though it can be made closed and time-like, it isn't everwhere geodesic.
}
}

\end{center}

What about non-radial paths?  Within the regular curve matching of Section~\ref{cvcont}, are there any closed piecewise geodesics in the DSC continuation that are at least partly timelike, i.e.\ timelike in one region and spacelike in the other?  It's easy to find a timelike path in one region that goes from $\Scrh^-$ to $\Scrh^+$.  Connecting to it, we'd also need a spacelike geodesic from a point on $r = 2 m$ out and back to $r = 2 m$.
In Appendix~\ref{ClosedPartTL} we examine the spacelike geodesic equation and show that a geodesic with the requisite properties does not exist.

With the alternative LK curve matching, timelike geodesics in {\rI} and {\rIII} can easily be paired up so that they form closed curves --- see Appendix~\ref{ClosedTLLK}.  Whether they are fully geodesic depends on how one views the kinks at the two junctions.

For the pita pocket with ``reverse'' junctions, curves start and end on the horizons, they do not pass through $\Sigma$, and thus there is no continuation difficulty.

\subsubsection{Discussion of Curve Joining Options}
\showlabel{cvcontcomp}

There are a couple of obvious arguments in favour of the LK continuation.  First, from all our experience in a regular Lorentzian spacetime, we expect that timelike worldlines should stay timelike.  Second, suppose the two spacetimes being joined were matter-filled, containing real matter derived from a well-behaved metric via the Einstein field equations, and not merely test matter.  In that case, there would definitely be matter flowing in a timelike direction on both sides of $\Sigma$, and it makes sense to just join up the two flows somehow.

On the down side, we would then have to explain how the matter flow in $M^-$ generates the matter flow in $M^+$, and the failure of momentum conservation becomes an important issue.  The cost of preserving the tardyonic character is therefore pretty extreme, and one wouldn't be able to portray the LK continuation as anything like normal Lorenzian behaviour.~%
\footnote{There is another rather tricky point.  If the two spacetime pieces really were matter-filled, they wouldn't be {\Schw}, and we wouldn't have nice simple null horizon surfaces to join them across.  While this does introduce some doubts about any {\Schw}-based conclusions, further consideration is well beyond the scope of this paper.}

The arguments in favour of the DSC continuation are, first, that it {\it seems\/} to preserve the inertial smoothness of Newton's laws, and, second, that the worldlines and curves agree better with how the coordinate lines and frame vectors are matched up.  If one accepts that Euclidean-to-Lorentzian signature change is a possibility, then DSC is a generalisation that does not introduce new problems.  Significant downsides are that signature change violates conservation laws anyway~\cite{HelDra94}, and also that two different families of ``worldlines'' change signature in two opposite senses, which is more complicated than all worldlines retaining their signatures.

It is apparent that one can't claim that either DSC or LK continuation is unambiguously more physical than the other, as there are problems with the physics in both cases.  One has to admit that with Euclidean ``spacetime'' and with signature change, it is not clear what the correct ``physical laws'' are, and there's no experiment to guide us.

\subsection{Structure at Infinity}
\showlabel{infinity}

The asymptotic structure of these spacetimes is somewhat peculiar.  Recall that $r$ is a valid affine parameter on (radial) Schwarzschild null geodesics, so that outgoing null geodesics start at the singularity and ``end'' on $\Scri^+$.  Similarly, $t-r^*$ is a valid affine parameter for the null geodesics that foliate $\Scri^+$, where $r^*$ is the usual ``tortoise'' coordinate.  Put differently, each copy of $\Scri^\pm$ in the maximally extended Schwarzschild spacetime is geodesically complete.

In the black mirror spacetime, however, we appear to be extending $\Scri^+$ from \rI\ to $\Scri^-$ from \rIII\ through the boundary ``points'' at $i^\pm$, see Figures~\ref{Flip1} and \ref{Pita1}.  Nonetheless, although null geodesics at finite radius that approach the horizon clearly reach it in finite affine ``time'', the corresponding ``parallel'' geodesics within $\Scri^+$ can not be continued similarly ``across the horizon'', at least, not in finite affine time.

We do not have a good interpretation of this behavior, other than to note that the structure ``at'' $i^\pm$ is quite complicated already in ordinary Schwarzschild spacetime.  Similar issues are present in the time-reversed case considered in Section~\ref{time} and Figures~\ref{AltMap} and \ref{Pita2}, in which two copies of $i^+$ are brought together at the future boundaries of two different copies of $\Scri^+$.

\section{Conclusions}
\showlabel{discuss}

We have examined the manifold underlying the black mirror construction presented by Tzanavaris, Boyle, and Turok in~\cite{Boyle25}, which involves joining the two exterior, $r \ge 2 m$, regions of the {\KrSz} spacetime along both their future and past event horizons.  We noted there are two causally different ways to do this matching, one where the time directions on each side continue ``forward'' at each junction, and one where they are ``reversed'' at both.  In the forward case, the timelike direction points towards the horizon on one side, and away from it on the other.  In the reverse case there are no timelike curves passing through; they just appear from the common past horizon, and end on the common future horizon, so the two parts are causally independent.

The black mirror spacetime~\eqref{black} as described in~\cite{Boyle25} is degenerate at the horizons, making it impossible to determine whether the horizons contain surface layers.  We have constructed the unique $C^1$ joining of the two exterior Schwarzschild regions across these horizons, and then investigated the possible presence of surface layers using several different methods, all of which agree.  There are none, and in this sense, we have confirmed the claim in~\cite{Boyle25} that the black mirror spacetime does not contain curvature singularities, although it is worth reiterating that our model and theirs use different differentiable structures at the horizons.  

However, a consequence of our forward model is that smooth timelike curves cannot remain timelike when crossing a horizon junction $\Sigma$, and conversely curves that stay timelike cannot remain smooth.  This is true both for timelike paths in $M^-$ being propagated through $\Sigma$ to the future, and for timelike paths in $M^+$ being traced back through $\Sigma$ to the past.  Furthermore, in the absence of a surface layer, there is no mechanism available to justify introducing a kink in such curves in order to preserve their timelike character.  This conundrum suggests that the black mirror spacetime may be unphysical, although without a $C^1$ differentiable structure it is difficult to establish the presence or absence of such kinks, or the presence or absence of a surface layer, in this model.

We also found that closed timelike piecewise geodesics are entirely possible with the Lorentzian kink continuation, but with the double signature change continuation there are no closed piecewise geodesics that are timelike in one of the joined regions.

Our results concerning this manifold construction, and its variants, may have relevance for the black hole information calculations based on it.

We end by collecting several issues whose resolution is beyond the scope of the present work.
First of all, the lack of a surface layer in the stress-energy tensor merely shows that there is no (nonzero) distributional term in the Einstein tensor, but is not sufficient to eliminate such terms in the full curvature tensor.  As analyzed in detail in~\cite{Mars93}, at a null junction there could still be (nonzero) distributional terms in the Weyl tensor.  However, we are not aware of any junction conditions at a null hypersurface with discontinuous metric that would permit the analysis of such terms.

Second, the differentiable structure at the intersection of the two horizons, the 2-sphere $U = 0 = V$, which involves a ``corner'' between two null 3-surfaces, is not adequately described by studying only its embedding into the separate (3-dimensional) horizons.

Thirdly, the ``reverse junctions'' mentioned in the latter half of section \ref{constrmirror}, present the problem of how to attach two distinct spacetime regions to a common black hole horizon (both in the future and in the past).  Is some kind of junction condition needed here, and might there be a surface layer?

Finally, the conformal structure of the black mirror spacetime deserves further analysis.  Are the two copies of $\Scri$, associated with \rI\ ($\Scri^+$) and \rIII ($\Scri^-$), distinct?  What is the structure at their apparent intersection, where the future timelike infinity ($i^+$) from \rI\ appears to meet the past timelike infinity ($i^-$) from \rIII?  In what sense, if any, does the black mirror spacetime contain a black hole?  Do the definitions of black holes and event horizons in terms of past boundaries of future timelike infinity need to be adjusted so as to better handle such cases?

\appendix

\section{Impossibility of Partly Timelike `Orbits' in DSC Continuations}
\showlabel{ClosedPartTL}

We here consider whether a piecewise geodesic with non-zero angular motion, that is timelike in {\rI} (say) and spacelike in {\rIII} can be made to close on itself.  We focus on the spacelike part, and show that if it is made to go out from the horizon at $r = 2 m$ and back, it cannot connect to a timelike direction across the junction.

It is far simpler to work with the geodesic equation in {\Schw} coordinates than in {\KrSz} coordinates.  We look at spacelike paths, $s = +1$, and stay in the $\phi = \pi/2$ plane, $w^a = (w^t, w^r, w^\theta, 0)$.  The spacelike and geodesic equations are
 \begin{align}
   s & = - \frac{(r - 2 m)}{r} \big( w^t \big)^2 + \frac{r}{(r - 2 m)} \big( w^r \big)^2 + r^2 \big( w^\theta \big)^2 ~,
 \showlabel{eq:SchwSpL} \\
   \ddot{t} = \frac{d^2 t}{d \alpha^2} & = - \frac{2 m}{r (r - 2 m)} \big( w^t \big) \big( w^r \big) ~,
 \showlabel{eq:GdEqT} \\
   \ddot{r} = \frac{d^2 r}{d \alpha^2} & = - \frac{(r - 2 m) m}{r^3} \big( w^t \big)^2 + \frac{m}{r (r - 2 m)} \big( w^r \big)^2 + (r - 2 m) \big( w^\theta \big)^2 ~,
 \showlabel{eq:GdEqR} \\
   \ddot{\theta} = \frac{d^2 \theta}{d \alpha^2} & = - \frac{2}{r} \big( w^r \big) \big( w^\theta \big) ~.
 \showlabel{eq:GdEqTh}
 \end{align}
Substituting from \eqref{eq:SchwSpL} into \eqref{eq:GdEqR} we have
 \begin{align}
   \big( w^t \big)^2 & = \frac{r^2}{(r - 2 m)^2} \big( w^r \big)^2 + \frac{r^3}{(r - 2 m)} \big( w^\theta \big)^2 - \frac{r}{(r - 2 m)} ~,
 \showlabel{eq:wT2SpL} \\
   \ddot{r} = \frac{d^2 r}{d \alpha^2} & = \frac{m}{r^2} - (3 m - r) \big( w^\theta \big)^2 ~,
 \showlabel{eq:GdEqRsb}
 \end{align}
and interestingly $w^r = \dot{r}$ vanishes from \eqref{eq:GdEqRsb}.  Clearly, $\ddot{r}$ can be negative if $r < 3 m$ and $w^\theta$ is big enough,
 \begin{align}
   \big( w^\theta \big)^2 > \frac{m}{r^2 (3 m - r)} ~.
 \showlabel{eq:wThMin}
 \end{align}
This value drops rapidly as $r$ decreases past $3 m$, and is minimum at $r = 2 m$.  (Such behaviour is the opposite of our intuition for timelike curves that become `straighter' as the angular speed increases.)  Indeed, there is a spacelike circular ``orbit'', $\ddot{r} = 0$,  $w^a = (0, 0, w^\theta, 0)$ at $r = 2 m$, $w^\theta = 1 / 2 m$.  

For a curve to emerge from $r = 2 m$, and then return into the same $r$ (actually going from {$\Scrh^-$} to {$\Scrh^+$}), there must be a point where $\dot{r} = w^r = 0$.  At this tangential point, \eqref{eq:wThMin} becomes an equality and the other geodesic equation components are
 \begin{align}
   \frac{d^2 t}{d \alpha^2} = 0 = \frac{d^2 \theta}{d \alpha^2} ~.
 \end{align}
How does the 4-velocity change from there?  Eq \eqref{eq:GdEqTh} shows that $w^\theta$ is speeding up where $w^r = \dot{r}$ is negative, while the minimum $|w^\theta|$ for negative $\ddot{r}$ is decreasing.  Therefore a spacelike geodesic that's tangential somewhere in $2 m < r < 3 m$, does bend inward and cross the horizon, both for extending the curve forward and backward from this point.

However, eq \eqref{eq:SplToSpl} shows that large $P^\theta$, and thus large $w^\theta$, is likely to preserve the spacelike character of a curve propagated through $\Sigma$.  Now transforming from $(U, V)$ to $(t, r)$,
 \begin{align}
   w^U \, w^V = \frac{e^{r/2m}}{32 m^3} \left\{ (r - 2 m) \big( w^t \big)^2 - \frac{r^2}{(r - 2 m)} \big( w^r \big)^2 \right\} ~,
 \end{align}
applying \eqref{eq:wT2SpL}
 \begin{align}
   w^U \, w^V = \frac{e^{r/2m}}{32 m^3} \left\{ r^3 \big( w^\theta \big)^2 - r \right\} ~,
 \end{align}
and going to the horizon, eq \eqref{eq:SplToSpl} --- in terms of $w^a$ rather than $P^a = \mu w^a$ --- becomes
 \begin{align}
   \big( w^\theta \big)^2 > \left| \big( w^\theta \big)^2 - \frac{1}{4 m^2} \right| ~,
 \end{align}
which, by \eqref{eq:wThMin} at $r = 2 m$, is always true.  In other words, spacelike geodesics which emerge from and return into the horizon always transition into spacelike paths at each $\Sigma$.  This means that timelike geodesics in one region cannot be closed off by spacelike geodesics in the other region.

\section{Timelike `Orbits' in LK Continuations}
\showlabel{ClosedTLLK}

In the case of LK curve continuations, it is very easy to connect timelike geodesics into closed curves.

We consider timelike radial geodesics, $s = -1$, $w^\theta = 0 = w^\phi$, for which \eqref{eq:SchwSpL} and \eqref{eq:GdEqR} give us $\ddot{r} = - m/r^2$.
In this case, $\ddot{r}$ is always negative, so if $\dot{r} = 0$ at any finite radius, the geodesic will go out from $r = 2 m$ and fall back in, emerging from $\Scrh^-$ and disappearing into $\Scrh^+$.  In the {\KrSz} diagram, the symmetries $(U,V) ~\leftrightarrow~ (V,U)$ and $(U,V) ~\leftrightarrow~ (-V,-U)$ ensure that if there's a timelike geodesic going from $(U, V) = (-A, 0)$ to $(U, V) = (0, B)$ in {\rI}, then there's one going from $(0, -A)$ to $(B, 0)$ in {\rIII}.  The endpoints of each match to the start points of the other, according to the LK prescription \eqref{LKcont}, thus completing a closed timelike curve.  Whether the acceleration of the curve is zero at the two junctions is up for discussion.  This can be generalised to the case of non-zero angular motion.

\end{document}